\documentclass[fleqn,usenatbib]{mnras}

\usepackage{newtxtext,newtxmath}
\usepackage{fontawesome} 
\usepackage{academicons} 

\usepackage[T1]{fontenc}

\usepackage{ulem}
\usepackage{hyperref}
\usepackage{color} 
\usepackage{multirow}
\usepackage{stfloats}

\usepackage[dvipsnames]{xcolor}

\newcommand{\be}{\begin{equation}}
\newcommand{\ee}{\end{equation}}
\newcommand{\ba}{\begin{eqnarray}}
\newcommand{\ea}{\end{eqnarray}}

\newcommand{\bi}{\begin{itemize}}
\newcommand{\ei}{\end{itemize}}
\newcommand{\mpch}{h^{-1} {\rm Mpc}}
\newcommand{\mpcht}{h^{-3} {\rm Mpc^3}}
\newcommand{\gpch}{h^{-1} {\rm Gpc}}
\newcommand{\gpcht}{h^{-3} {\rm Gpc^3}}
\newcommand{\hmpc}{h {\rm Mpc}^{-1}}

\newcommand{\hgpc}{h{\rm Gpc}^{-1} }

\newcommand{\Msun}{M_{\odot}}
\newcommand{\dif}{\mathrm{d}}

\newcommand{\nside}{N_\mathrm{side}}

\newcommand{\camb}{{\texttt{CAMB}}}

\newcommand{\degree}{^{\circ}}

\newcommand{\Rmnum}[1]{\uppercase\expandafter{\romannumeral #1}}
\DeclareRobustCommand{\VAN}[3]{#2}
\let\VANthebibliography\thebibliography
\def\thebibliography{\DeclareRobustCommand{\VAN}[3]{##3}\VANthebibliography}

\usepackage{graphicx}	
\usepackage{amsmath}	
\usepackage{amssymb}	

\title{Box Replication Effects in Weak Lensing Light-cone Construction}

\author[Chen et al.]{
Zhao Chen,$^{1,2}$
Yu Yu,$^{1,2}$\thanks{E-mail: yuyu22@sjtu.edu.cn}
\\
$^{1}$Department of Astronomy, Shanghai Jiao Tong University,
800 Dongchuan Road, Shanghai 200240, China\\
$^{2}$Key Laboratory for Particle Astrophysics and Cosmology (MOE)/Shanghai Key Laboratory for Particle Physics and Cosmology, Shanghai 200240, China\\\
}

\date{Accepted XXX. Received YYY; in original form ZZZ}

\pubyear{2024}

\begin{document}
\label{firstpage}
\pagerange{\pageref{firstpage}--\pageref{lastpage}}
\maketitle

\begin{abstract}
Weak gravitational lensing simulations serve as indispensable tools for obtaining precise cosmological constraints. 
In particular, it is crucial to address the systematic uncertainties in theoretical predictions, given the rapid increase in galaxy numbers and the reduction in observational noise. 
Both on-the-fly and post-processing methods for constructing lensing light-cones encounter limitations due to the finite simulated volume, necessitating the replication of the simulation box to encompass the volume to high redshifts.
To address this issue, our primary focus lies on investigating and quantifying the impact of box replication on the convergence power spectrum and higher-order moments of lensing fields. 
Subsequently, a univariate model is utilized to estimate the amplitude parameter $A$ by fitting four statistics measured from partial-sky light-cones along specific angles, to the averaged result from random directions.
The investigation demonstrates that the systematic bias stemming from the box replication phenomenon falls within the bounds of statistical errors for the majority of cases.
However, caution should be exercised when considering high-order statistics on a small sky coverage ($\lesssim 25~\mathrm{deg^2}$).
For this case, we have developed a code that facilitates the identification of optimal viewing angles for the light-cone construction. 
This code has been made publicly accessible 
\href{https://github.com/czymh/losf}{\faGithub}.

\end{abstract}


\section{introduction}
\label{sec:introduction}
With the advent of cosmic microwave background radiation (CMB) observations, we have entered an era of precision cosmology.
It is now widely recognized that dark matter and dark energy comprise the primary constituents of our Universe. 
In order to delve into the essence of these enigmatic components, it is imperative to meticulously measure the geometry and expansion history of the cosmos.
The current Stage-\Rmnum{3} dark energy experiments include the Dark Energy Survey (DES\footnote{\url{http://www.darkenergysurvey.org/}}), the Hyper Suprime-Cam Subaru Strategic Program (HSC\footnote{\url{https://www.naoj.org/Projects/HSC/}}), the Kilo-Degree Survey (KiDS\footnote{\url{http://kids.strw.leidenuniv.nl/}}) and the Extended Baryon Oscillation Spectroscopic Survey (eBOSS\footnote{\url{https://www.sdss.org/surveys/eboss/}}) which contribute to the large precise data acquisition.
The pipelines of data analysis developed in these surveys consider nonlinear theoretical prediction, complicated observational noise, and significant systematic error.
As a result, these surveys can obtain accurate and precise parameter constraints for the standard $\Lambda$CDM model, and even beyond.
The next generation of sky surveys such as the Dark Energy Spectroscopic Instrument (DESI\footnote{\url{https://www.desi.lbl.gov/}}), the Vera Rubin Observatory Legacy Survey of Space and Time (LSST\footnote{\url{http://www.lsst.org}}), Euclid\footnote{\url{http://www.euclid-ec.org}} and the China Space Station Telescope (CSST\footnote{\url{https://www.nao.cas.cn/csst/}}) will constrain the cosmological parameters to an unprecedented precision and may extend our knowledge about dark energy and dark matter.

Weak gravitational lensing or cosmic shear, which is the light deflection effect caused by the gravitational potential along the line of sight, is a promising cosmological probe to measure dark matter distribution directly \citep[e.g.,][]{2001PhR...340..291B,2015RPPh...78h6901K,2018ARA&A..56..393M}.
Thanks to the ongoing and upcoming wide-field imaging surveys, a large number of observational data can help us to tighten cosmological constraints power, especially for mean matter density $\Omega_m$, mass density fluctuation within a sphere of $R = 8~\mpch$ today $\sigma_8$ and the equation of state of dark energy $w_0$ \citep[e.g.,][]{2005astro.ph.10266M, 2017MNRAS.465.1454H, 2018PhRvD..98d3528T, 2020PASJ...72...16H}.
The reduction of confidence intervals needs both precision observation and accurate theoretical predictions \citep[e.g.,][]{2023MNRAS.tmp.3423Y}.
Due to the non-linear evolution in the late Universe, linear perturbation theory fails at small scales.
To fully explore the cosmological information of data, it is essential to generate accurate numerical simulations for weak lensing surveys \citep[e.g.,][]{2000ApJ...530..547J,2009A&A...499...31H,2018MNRAS.481.2813G,2018ApJ...853...25W}.

In cosmological simulations, both dark matter and baryonic matter are usually treated as particles and their evolution is simulated within a cubic box.
The output of these simulations consists of the positions and velocities of the structures at different time intervals, often referred to as \textit{snapshots}.
These snapshots represent discrete time slices of the simulated universe, allowing researchers to study the evolution of cosmic structures over cosmic time.
In the process of combining observations along a null geodesic with simulations, it is necessary to transform the coordinates of the simulated data to a light-cone geometry. 
For a more realistic mock catalog for galaxy clustering analysis, this transformation is required to properly account for various observational effects, such as selection effects, projection effects, and line-of-sight contamination \citep[e.g.,][]{2019A&A...631A..82I,2022LRCA....8....1A}.
Previous works have employed different methods to construct a light-cone from simulation data, with one common approach being post-processing adjacent snapshots and their corresponding haloes.
In the simplest case, to fill in the shell between comoving radial distance $\bar\chi-\frac{1}{2}\Delta\chi$ and $\bar\chi+\frac{1}{2}\Delta\chi$, we directly pick out the structure from a snapshot whose redshift is closest to $z(\bar\chi)$.
This method can provide accurate statistical predictions when the simulation outputs are densely sampled.
However, it is important to note that discontinuities can arise at the boundaries of the snapshots, which may introduce small-scale systematic errors, particularly when the number of snapshots is insufficient to adequately capture the desired spatial and temporal resolution.
To mitigate these systematic errors, researchers employ an interpolation process on the phase-space coordinates of particles, haloes, and subhaloes \citep[e.g.,][]{2013MNRAS.429..556M,2017MNRAS.470.4646S,2019A&A...631A..82I}.

Simulating the high-redshift survey for weak lensing is challenging due to several factors.
On the one hand, lensing simulations need large sky coverage to match the wide-field survey and suppress the super-sample covariance effects \citep[e.g.,][]{Harnois-Deraps:2015aa,2020ApJ...897...14C,2021JCAP...02..047S}.
On the other hand, to accurately model the evolution of subhaloes within the halo light-cones, underlying simulations with sufficient resolution and physics are necessary \citep[e.g.,][]{2020MNRAS.493..305H}.
Hence, precise lensing mocks need both a large simulated volume and a substantial quantity of particles, which is computationally impossible for high redshift surveys.
To extend the maximum redshift of a light-cone simulation while balancing computational resources and precision, one approach is to take advantage of the periodicity of the simulation box by replicating it along its preferred axis.
However, this replication can introduce artificial structures known as the \textit{box replication effect}, which manifests as a kaleidoscope-like pattern in the lensing map when the simulated box is small or the redshift is relatively high \citep[e.g.,][]{2005MNRAS.360..159B,2007MNRAS.376....2K}.
To mitigate this effect, researchers have explored various strategies.
One option is to select some special viewing angles to avoid too many times of structure repetition \citep[e.g.,][]{2007MNRAS.376....2K,2009A&A...499...31H,2010ApJS..190..311C,2016ApJS..223....9B}.
This method only works well when the opening angle of a pencil-beam light-cone is small enough.
Another approach is to avoid the excessive repetition of structures by randomly shifting and rotating the shells or shell-bundles so that at different redshifts the volumes are uncorrelated on the concerned large scales \citep[e.g., ][]{2005MNRAS.360..159B,2016MNRAS.460.1173R,2019ApJS..245...26K, 2024MNRAS.530.5030O}.
The above randomisation is also employed to generate multiple pseudo-independent realizations. 
It indeed improves the estimation of the covariance for various lensing statistics by assuming the statistical independence of these realizations \citep[e.g.,][]{2019JCAP...01..044S,2021JCAP...02..047S}.
Unfortunately, \citealt{2024MNRAS.529.1862U} confirm that the rotation scheme introduces an underestimation for lensing power spectrum by using the recent state-of-art \textbf{FLAMINGO} simulations with large boxes ($5.6\ \mathrm{Gpc}, 11.2\ \mathrm{Gpc}$).
The randomisation across different 12 shell-bundles introduces a suppression ($\sim 2\%$) at $\ell \sim 100$ for the CMB lensing convergence power spectrum.
The underestimation can reach even $20\%$ if we rotate 72 shells individually in the convergence field construction.

The on-the-fly light-cone creation technique is another approach used to construct light-cones in simulations, and it offers advantages over post-processing methods \citep[e.g.,][]{2002ApJ...573....7E,2008MNRAS.391..435F,2017ComAC...4....2P,2021MNRAS.506.2871S}.
In this technique, particles are recorded only when they intersect the backward light-cone in the simulation coordinate system.
The corresponding halo and subhalo catalogs can also be outputted on the fly.
Therefore, there are not any ambiguous and discontinuous structures for the on-the-fly light-cones compared to post-processing constructions.
While it is also limited by the finite volume of the underlying N-body simulation.
Periodic replication of the simulation volume is required to extend the past light-cone to a high redshift.
In principle, there is the same box replication effect for both the post-processing method and on-the-fly light-cone construction.

In this paper, we focus on the exploration and quantification of the box replication effect on lensing statistics.
The investigation is based on the post-processing of simulation snapshots.
In principle, this study can be extended to the box replication effect for the on-the-fly construction.
This paper is organized as follows. In section \ref{sec:theory}, we provide a concise overview of the knowledge pertaining to weak gravitational lensing and the statistical tools associated with it.
Section \ref{sec:simulations} outlines the utilization of N-body simulations and post-processing techniques for constructing a particle-based light-cone.
The examination of the box replication effect on the power spectrum and higher-order convergence moments, and the influence on parameter fitting are investigated in Section \ref{sec:results}.
Moreover, we put forth a methodological framework for the determination of optimal viewing angles to construct the light-cone to avoid this artificial effect.
Subsequently, an efficient pipeline is proposed to provide optimal line-of-sight directions for light-cone construction.
Finally, in Section \ref{sec:conclusion}, we summarize and draw conclusions based on our key findings.

\section{Lensing Theory}
\label{sec:theory}
In this section, we present a brief introduction to the theoretical background.
We first review the basic knowledge of weak gravitational lensing, then the commonly used convergence power spectrum and high-order statistics.

\subsection{ weak gravitational lensing}

Weak gravitational lensing is caused by the whole components of matter, which provides the most potential probe to the hiding dark matter.
For a source galaxy with the real position $\boldsymbol{\beta}$ and observed position $\boldsymbol{\theta}$, we can define the lens equation as 
\be
\boldsymbol{\beta} = \boldsymbol{\theta} - \boldsymbol{\alpha}\ ,
\label{eq:lenseq}
\ee
where $\boldsymbol{\alpha}$ is the so-called deflection angle.
In general, the distortion from image coordinates $\boldsymbol{\theta}$ to source coordinates $\boldsymbol{\beta}$
can be characterized by the Jacobi matrix $\mathcal{A}=\frac{\partial \boldsymbol{\beta}}{\partial \boldsymbol{\theta}}$.
In linear order, the matrix components are determined by the second derivatives of the gravitational potential
\be
\begin{aligned}
A_{i j} &=\frac{\partial \beta_{i}}{\partial \theta_{j}}=\delta_{i j}-\frac{\partial \alpha_{i}}{\partial \theta_{j}} \\
&=\delta_{i j}-\frac{2}{c^{2}} \int_{0}^{\chi} \mathrm{d} \chi^{\prime} \frac{\left(\chi-\chi^{\prime}\right) \chi^{\prime}}{\chi} \Phi_{, i j}\left(\chi^{\prime}\boldsymbol{\theta}, \chi^{\prime}\right)\ .
\end{aligned}
\label{eq:jacobi}
\ee
Here, $\Phi$ is the gravitational potential, and $\chi$ is the comoving angular distance.
In the weak lensing study, we define the symmetrical matrix $\mathcal{A}$ as the following form
\be
\mathcal{A}=\left(\begin{array}{cc}
1-\kappa-\gamma_{1} & -\gamma_{2} \\
-\gamma_{2} & 1-\kappa+\gamma_{1}
\end{array}\right)\ ,
\label{eq:jacobi2}
\ee
where convergence $\kappa$ describes the isotropic increase or decrease of the intrinsic galaxy size and cosmic shear $\gamma = \gamma_1 + i\gamma_2$ represents the shape distortion in different directions.
Using the Poisson equation, the scalar component $\kappa$ can be written as the weighted integral of the over-density $\delta$ along the line of slight,
\be
\kappa\left(\hat{n}, \chi_{s}\right)=\int_{0}^{\chi_{s}} \frac{3 \Omega_{m} H_{0}^{2}}{2 c^{2}}\frac{\chi \left(\chi_{s}-\chi\right)}{\chi_{s}} \delta(\hat{n}, \chi) \frac{d \chi}{a\left(\chi\right)}\ .
\label{eq:kappa}
\ee
Here, $H_0$ is the Hubble constant, $\Omega_m$ is the matter density fraction at present time, $c$ is the speed of light, and $a(\chi)$ is the scale factor at $\chi$.
The lensing kernel is defined as
\be
W\left(\chi, \chi_{\mathrm{s}}\right)=\frac{3 \Omega_{m} H_{0}^{2}}{2 c^{2}}(1+z) \frac{\chi \left(\chi_{\mathrm{s}}-\chi\right)}{\chi_{\mathrm{s}}}\ ,
\label{eq:lensingkernel}
\ee
where $\chi_s$ is the comoving distance at the source redshift $z_s$.
Considering the slow change of the lensing kernel, the integral reduces into summation under the Born approximation \citep{Cooray:2002aa}.
The convergence map can be easily obtained by the linear summation of projected dark matter density fields $\delta^{\Sigma}_{i}(\hat{n})$.
\be
\kappa(\hat{n})=\sum_{i} W_{i} \delta_{i}^{\Sigma} \Delta \chi\ ,
\label{eq:born}
\ee
where $W_i$ is the lensing kernel for the $i$-th density shell and $\Delta\chi$ is the thickness of lens planes.

In this work, we fix the thickness of each lens plane at $100~\mpch$.
Because in \citealt{2020AJ....159..284Z} two light-cone parameters affecting the accuracy of Gaussian or non-Gaussian statistics have been investigated and they found that using the thin shell ($\Delta \chi<60~\mpch$) leads to the underestimation of the power spectra at all scales.
Additionally, \citealt{2020MNRAS.493..305H} discussed that the systematic error due to Born approximation on the weak lensing power spectra is below $1 \%$ level for the angular scales $\ell = 100$ to $10000$.  
In this paper, we generate the convergence maps by using Eq.~\ref{eq:born} instead of the time-consuming ray-tracing technology and consider that the impact of the first-order Born approximation is negligible.

\subsection{ Statistics }
\label{stats}

\subsubsection{ Convergence Power Spectrum }
The two-point correlation function is one of the most commonly used statistics to quantify the weak lensing signal.
Its counterpart, the power spectrum measures the convergence fluctuations over different angular scales, which is related to the projected matter density along the line of sight.
In Fourier space, we can easily obtain this second-order statistic from the numerical simulations
\begin{equation}
    \left\langle\hat{\kappa}(\boldsymbol{\ell}) \hat{\kappa}^*\left(\boldsymbol{\ell}^{\prime}\right)\right\rangle=(2 \pi)^2 \delta_{\mathrm{D}}\left(\boldsymbol{\ell}-\boldsymbol{\ell}^{\prime}\right) C^{\kappa\kappa}(\ell)\ ,
    \label{eq:ckk-measure}
\end{equation}
where $\delta_D$ is the Dirac-$\delta$ function.
Theoretically, the convergence power spectrum can be computed from the three-dimensional matter power spectrum by using a projection formula that involves a lensing kernel and a Limber approximation
\be
C^{\kappa \kappa}(\ell)=\int_{0}^{\chi_{\mathrm{s}}} \mathrm{d} \chi \frac{W(\chi)^{2}}{\chi^{2}} P_{\delta}\left(k=\frac{\ell}{\chi}\ , z(\chi)\right),
\label{eq:ckappa}
\ee
where the $P_{\delta}$ is the non-linear matter power spectra at the given redshift $z(\chi)$.
The convergence power spectrum is sensitive to both the amplitude and the shape of the matter power spectrum, as well as the growth of structure and the geometry of the Universe. 
Therefore, accurate prediction is essential to obtain valuable cosmological constraints from lensing surveys.
However, perturbation theory fails at small scales (large $\ell$) because of the nonlinear evolution in the late Universe.
For the lensing two-point statistic, we could use high-resolution N-body simulation boxes to accurately measure the matter power spectrum $P_\delta(k)$ at different redshifts, then use the Limber integral to predict lensing signals.
However, for the statistics beyond two-point which are useful to break the parameter degeneracy, the post-processing weak lensing simulations or on-the-fly light-cones are essential.

\subsubsection{ Convergence Moments }

The second moment is related to the power spectrum and the variance of convergence fields.
The higher-order moments measure the non-Gaussianity.
They can capture additional cosmological information to provide tighter constraints on $\Omega_m$ and $\sigma_8$ (e.g.,  \citealt{2015PhRvD..91j3511P, 2020MNRAS.498.4060G,2022PhRvD.106h3509G}).
In this paper, we investigate the box replication effect on the second, third, and fourth convergence moments.

Moments of weak lensing fields are commonly expressed as $\left\langle\kappa^n_{\theta}\right\rangle$ that are a function of smoothing scales $\theta$.
We use the simplest estimators to measure the convergence moments of smoothed maps
\begin{equation}
\left\langle \kappa^{n}_{\theta} \right\rangle =\frac{1}{N_{\mathrm{tot }}} \sum_\mathrm{pix}^{N_{\mathrm{tot}}} \kappa_{\theta, \mathrm{pix}}^n\ ,
\end{equation}
where $N_\mathrm{tot}$ represents the total number of pixels for a given convergence map.
Here, we choose 15 smoothing scales inside an interval $\theta \in [3.2, 100]~\mathrm{arcmin}$ and use equally log-spaced bins.
It is worth noting that our work does not provide theoretical formulas for convergence moments, as our main focus is on the changes caused by the box replication effect on this statistic.
We recommend interested readers to \citealt{2020MNRAS.498.4060G}.


\section{datasets}
\label{sec:simulations}

In this section, we describe the numerical simulations and the details of our light-cone construction.
Note that we mainly focus on the box replication effect, and other high-order systematics are not investigated in this work.
We do not use the ray-tracing technique to obtain weak lensing maps and adopt Born approximation to simplify the calculations.

\subsection{CosmicGrowth}
\label{sec:cosmicgrowth}

For cosmological numerical simulations, the box size and the number of particles are two crucial parameters. 
In a simulation with a fixed side length $L$, the cosmological information at scales $k < 2\pi/L$ vanishes. 
The missing of long-wave fluctuations for a limited simulation box leads to systematic errors for both the power spectrum and its covariance. 
Similarly, the dark matter halos with very low mass cannot be adequately resolved for a deficient number of particles, which can not satisfy the requirement of the current emission line galaxy (ELG) survey \citep[e.g.,][]{2023ApJ...954..207G}.
Several works have confirmed that the simulated volume $V \gtrsim 1\ \gpcht$ and a mass resolution $M_\mathrm{p} \sim 10^9\ h^{-1} \Msun$ are required for the Stage-\Rmnum{4} surveys to ensure the percent-level convergence of different measurements and statistical errors \citep[e.g.,][]{2016JCAP...04..047S,2019MNRAS.489.1684K}.

Therefore, we use WMAP\_3072\_1200 (hereafter L1200), one of the high-resolution simulations in the state-of-the-art N-body simulation suite \textit{CosmicGrowth} to construct the lensing light-cone.
The simulation configuration is detailed in Tab. 4 in \citealt{Jing:2019uw}. 
The mass of each particle is about $4 \times 10^9\ h^{-1}\mathrm{M_{\odot} }$.
The cosmological parameters are from WMAP results  \citep[e.g.,][]{2011ApJS..192...18K, 2013ApJS..208...19H}, with $\Omega_c = 0.2235, \ \Omega_b = 0.0445, \ \sigma_8=0.83, \ n_s=0.968$ and the dimensionless Hubble constant $h=0.71$.
The initial redshift $z_i = 144$ and 100 snapshots are output between $z=16.87$ and $z=0.0$ at the equal logarithm space of scale factor.
The box length of this simulation is $1.2\ \hgpc$ on each side.
We only use the one-eighth particles to construct the particle light-cones considering the computational cost.
Matter and convergence power spectra have been checked at different redshifts.
We find excellent agreements between full sample and sub-sample at even non-linear regions.

\subsection{ Light-cone Construction }
\label{sec:lightcone}
\subsubsection{ Full-sky Lignt-cone }
\label{sec:full-sky-lc}

\begin{figure}
	\includegraphics[width=0.9\columnwidth]{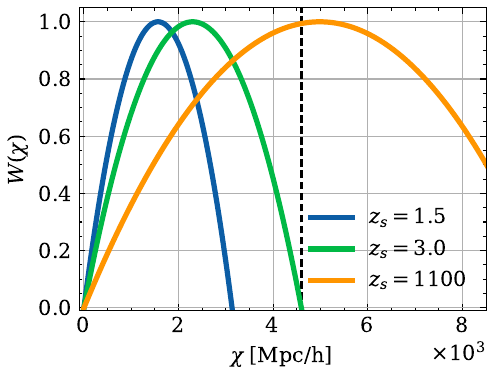}
    \caption{The lensing kernels of weak lensing convergence for three source redshifts. }
    \label{fig:Wkappa}
\end{figure}

Our aim is to investigate the box replication effect for different line-of-sight directions spanning the whole sky.
To avoid too many repeats of the construction process for each line-of-sight, we choose to obtain the full-sky maps first.
Simulation boxes at different redshifts are directly tilled together along three main axes to cover the past light-cone geometry from $z=0$ to $z_s$.
Besides, we also construct the corresponding CMB lensing map by using the same density fields for each snapshot to investigate the effects of different lensing kernels.
In this case, the contribution to convergence power from redshift higher than $z_s = 3.0$ is set to zero by construction.
In Fig.~\ref{fig:Wkappa}, the lensing kernels at different source redshifts are shown by different colors, and the dashed line is located at comoving distance $\chi(z_s=3.0) = 4.6\ \gpch$. 

Then each light-cone is made up of a set of discrete projected spherical density shells which are calculated by HEALPix\footnote{\url{https://healpy.readthedocs.io/en/latest}}\citep{2005ApJ...622..759G}.
We projected the dark matter particles to each shell at a comoving distance $\chi$ using the snapshot with the nearest redshift ($z_i \sim z(\chi)$) to capture the evolution of redshift.
According to \citealt{2020AJ....159..284Z}, it is reasonable to fix projection depth at $100~\hmpc$ for weak lensing survey.
The Healpix parameter $N_\mathrm{side} = 8192$, and the corresponding pixel resolution $\Delta \theta=\sqrt{3 / \pi} / N_{\text {side }} = 0.43~\mathrm{arcmin}$.

Similar to \citealt{2021JCAP...02..047S}, the convergence located in pixel $\theta_\mathrm{pix}$ is given by the weighted summation of all lens planes along the line of sight
\be
\kappa\left(\theta_{\mathrm{pix}}\right) \approx \frac{3\Omega_{m}H_0^2}{2c^2}  \sum_{i} W_{i} \frac{N_{\mathrm{pix}}}{4 \pi} \frac{V_{\text {sim }}}{N_{\mathrm{p}}} \frac{n_{p}\left(\theta_{\mathrm{pix}}, \Delta \chi\right)}{\chi^{2}\left(z_{i}\right)}\ ,
\label{eq:kappa_pix}
\ee
where $N_{\mathrm{pix}}$ is the total number of pixels on the whole sky, $V_{\text {sim }}$ is the volume of the simulation box, and $N_\mathrm{p}$ is the total particle number in a cubic snapshot.
For the $i$-th lens plane, the projected number density $n_{p}\left(\theta_{\mathrm{pix}}, \Delta \chi_{i}\right)$ should be weighted by the lensing kernel $W_i(\chi_i, \chi_s)$.
Here, $\chi_i = (2i-1)\Delta\chi/2$ and $\chi_s = \chi(z_s)$ are the comoving angular diameter distance of $i$-th lens plane and source plane, respectively.
It is crucial to emphasize that no randomization procedure is incorporated in the light-cone construction in this work. 
Nevertheless, in Appendix~\ref{app:random}, we additionally investigate two rotation strategies intended to mitigate repetitive structures. 

In order to suppress the cosmic variance, we chose 25 different observers for the full-sky light-cone creation.
All observer positions are generated by using Sobol sequence \citep{SOBOL196786} to obey the Copernican principle.

\subsubsection{ Partial Light-cone }
\label{sec:partial-lc}

To isolate the replication effects across various sky coverage areas, it is necessary to subdivide the full-sky convergence map into multiple sub-maps.
Each sub-map should be centred along a distinct line-of-sight direction.
To facilitate comparison, we first need to define the special and normal line-of-sight directions.

Based on the geometry of tiling, we can categorize three distinct types of special directions.
They are the main axis directions (comprising 6 vectors denoted as $(1,0,0) \times 6$), the diagonal directions of the x-y/y-z/x-z planes (including 12 viewing angles notated as $(1, 1, 0) \times 12$), and the diagonal orientations of cubic boxes (containing 8 diagonal vectors notated as $(1, 1, 1) \times 8$). 
Since the identical large-scale structure manifests repeatedly along these three unique direction types, sub-maps aligned with these directions inherently exhibit notable box replication effects.
To obtain a reliable conclusion, it is essential to include random lines of sight to serve as a baseline for comparison. 
We achieve this by uniformly dividing the entire celestial sphere into $12\nside^2$ directions, employing the \texttt{ipix2vec} function available in the HEALPix framework. 
For this study, we have chosen a value of $\nside = 4$, resulting in $192$ directions and associated sub-maps that serve as validation samples.
Note that none of the 192 random directions overlaps with the three kinds of special directions.
We call these validation samples as random light-of-sight samples although they are approximately uniformly distributed on the sphere.
Additionally, it is essential to mention that the rotation of shell or shell-bundles is not employed to remove the box replication effect since this process also changes the statistics slightly, as detailed in Appendix~\ref{app:random}.

In practice, the full-sky convergence map is rotated so that a specified line of sight is located at a specific angular position, i.e., $\hat{n} = (\theta=0\degree, \phi=0\degree) = (1,0,0)$.
This is to avoid the high latitude projection effect from sphere to flat plane, and the equatorial zone can be pixelized to regular diamonds with minimal shape distortion by the HEALPix scheme.
It is important to note that the rotation is performed in the spherical harmonic space to ensure the preservation of the lensing signal before and after rotation.
Subsequently, we can extract the lensing signals in the vicinity of this particular direction using a square mask with an area of $5\times 5~\ \mathrm{deg^2}$ as the fiducial case.
The significance of the box replication effect is obviously dependent on the mask size. 
Thus, the different choices of mask size in this post-processing operation are also investigated in Sec.~\ref{sec:mask}.
We decrease the map resolution from $0.43^{\prime\prime}$ to $3.44^{\prime\prime}$ for a balance of speed and accuracy in the process.


\section{Results}
\label{sec:results}
In this section, we show the effects of the box-replication on the two-point statistics and high-order statistics.

\subsection{Validation of Full-sky Convergence}

\begin{figure}
    \centering
    \includegraphics[width=0.45\textwidth]{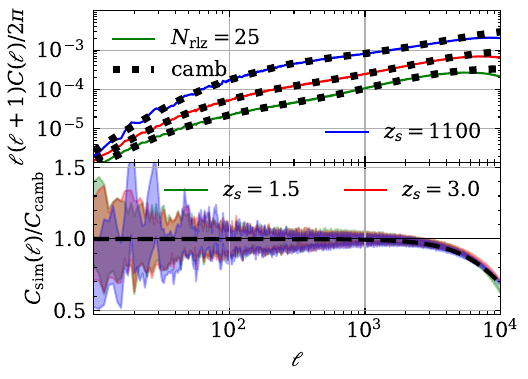}
    \caption{\textit{Top}: Validation of full-sky convergence power spectrum.
    The red and blue lines are from 25 realizations using Born approximation, while the thick dotted lines are the $\camb$ theoretical predictions from Limber integration.
    \textit{Bottom}: We show the ratio between measurements and theoretical predictions. 
    The colour regions indicate the standard deviation of 25 different observers.
    The black dashed line is determined using the formula outlined in Eq. 5 of \citealt{2017ApJ...850...24T} by setting $\ell_\mathrm{res} = 1.8 \nside$ to describe the effect from limited resolution.
    }
    \label{fig:full-sky-ps}
\end{figure}

Power spectrum and its counterpart, the two-point correlation function are widely used in cosmological analysis.
First of all, we present the power spectra of the full-sky convergence maps in the top panel of Fig.~\ref{fig:full-sky-ps}.
The red and blue lines represent the dimensionless power spectrum of weak lensing and CMB lensing, respectively.
The black dotted lines indicate the results of Eq.~\ref{eq:ckappa}, where the underlying density power spectrum is calculated by using $\camb$ \footnote{\url{https://camb.readthedocs.io/en/latest/}},
which integrates the enhanced halo-model fitting formula, \texttt{HMCODE}, to offer accurate nonlinear matter power spectrum predictions across various redshifts \citep[][]{2015MNRAS.454.1958M}.
To aid comparison, we display the ratio between simulation results and theoretical predictions in the bottom panel.
The colored region represents the standard deviation of the measurements from 25 realizations.

Our analysis reveals that the measured two-point statistics closely match the theoretical predictions for $\ell \lesssim 2000$ across different source redshifts.
At large scales (small $\ell$), the large scatter is caused by the sample variance.
The remarkable agreement between the measured statistics and the theoretical predictions highlights the effectiveness of the Born approximation, a finding consistent with previous studies \citep[e.g.,][]{2018ApJ...853...25W, 2020MNRAS.493..305H}.
The discrepancy observed between theoretical predictions and simulation results at angular scales $\ell \gtrsim 2000$ primarily stems from the limited resolution of HEALPix maps. 
This effect can be approximately characterized by the equation provided in \citealt{2017ApJ...850...24T}:
\begin{equation}
C_{\kappa\kappa}(\ell) \rightarrow \frac{C_{\kappa\kappa}(\ell)}{1 + (\ell/\ell_\mathrm{res})^2}, 
\label{eq:powersupression}
\end{equation}
where $\ell_\mathrm{res} = 1.8 \nside$.
This finding is consistent with the results of previous papers \citep[e.g.,][]{2023MNRAS.524.5591F}.

\subsection{ Partial-sky Convergence Power Spectrum }

\begin{figure}
    \centering
    \includegraphics[width=0.45\textwidth]{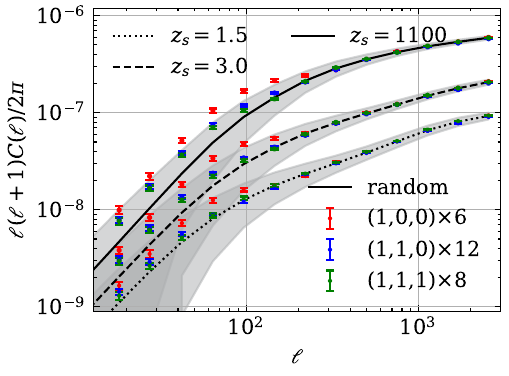}
    \caption{ Convergence power spectrum of light-cones with an area of $5 \times 5~\mathrm{deg}^2 $ along random line-of-sight directions (solid black lines) and three special kinds of orientations (data points).
    The shaded region indicates the standard deviations of $192$ masked maps while the data points and the error bars show the measurement errors of the power spectrum for special orientations.
    }
    \label{fig:masked-ps}
\end{figure}

In this analysis, we compare the partial-sky convergence power spectra of random samples with three distinct special directions.
Here, the ``partial-sky" means that we use a mask to highlight the presence of repeated structures.
Because these artificial structures faded away quickly as the line of sight deviated from a particular direction.
Similar to the previous figure, Fig.~\ref{fig:masked-ps} shows the power spectrum of the masked convergence field for three different source redshifts, 1.5 (dotted), 3.0 (dashed) and 1100 (solid), obtained from the L1200 simulation.
The black lines represent the computed mean spectra from $25 \times 192$ realizations of partial light-cones along random viewing angles.
In Appendix~\ref{sec:app-random} we demonstrate that employing the statistics from the aforementioned 192 random directions as a reference is appropriate, as the statistics from 768 directions and those from 128 safe directions with no-repetition (as discussed in later section \ref{sec:losf}) is consistent with each other.
Shaded regions are utilized to indicate the statistical error of the spectra.
The results of three different types of special samples are presented in red, blue, and green data points,
accompanied by the error bars depicting the measurement errors.

Compared to the random samples, the first kind of special directions (red points) shows noticeable power spectrum excess at $\ell \lesssim 200$. 
Smaller overestimations can also be found for two other special directions (the blue and green data points).
This could be attributed to the fact that the number of structure replications along these directions is less than the $(1,0,0)\times6$ directions.
It is worth noting that the overestimated signals are observed for different source redshifts, suggesting that this effect is not strongly influenced by the specific kernel but rather is predominantly determined by the number of structure repetitions.

\begin{figure*}
    \centering
    \includegraphics[width=0.96\textwidth]{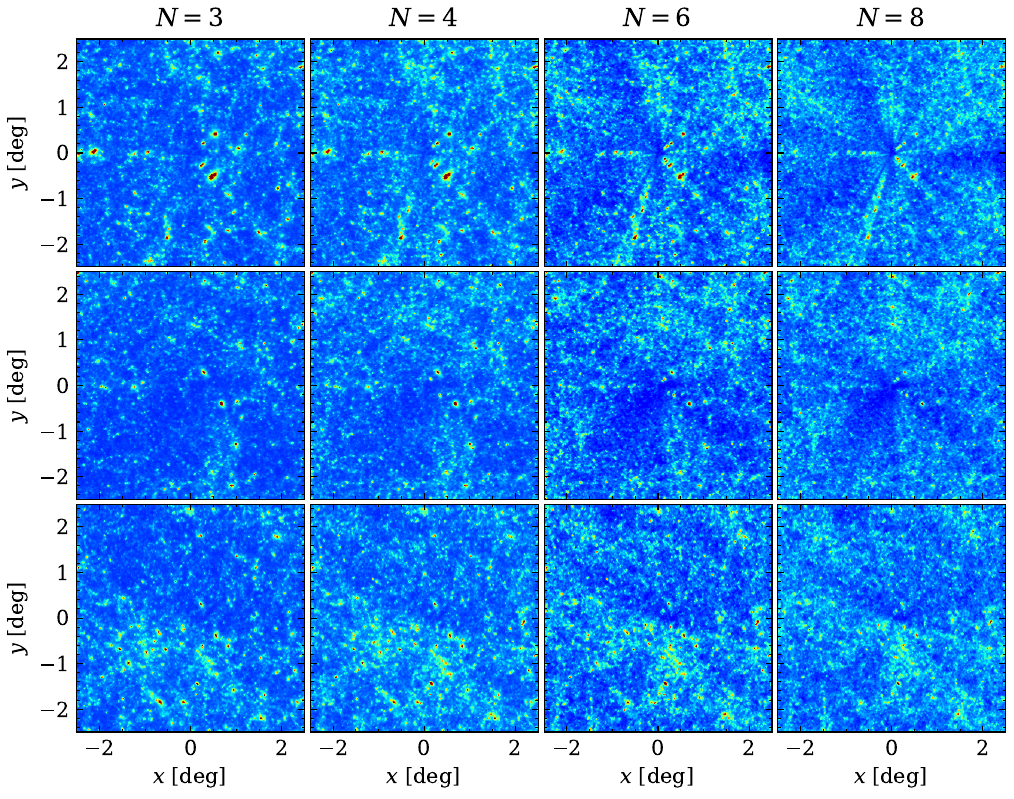}
    \caption{    
    Visualization of partial-sky convergence map of $5\times5~\mathrm{deg^2}$ using $L=250~\mpch$ simulation boxes.
    Different box repetition times $N$ also indicate the distance of the source plane to the observer, $\chi_s=250N~\mpch$.
    The stacking of partial-sky light-cones along the most special direction $(1,0,0)$ is shown in the upper panels.
    The middle and lower panels represent the sky area with the same size as the upper panels but for the line-of-sight direction $(1, 1, 0)$ and $(1,1,1)$, respectively.
    Note that for our case using L1200 simulation, $N \approx 2.6,\ 3.8 $ for $z_s = 1.5,\ 3.0$, respectively.
    }
    \label{fig:visual}
\end{figure*}

To further support our conclusion, we utilized the well-known cosmological simulation code, Gadget-4 (as described in \citealt{2021MNRAS.506.2871S}), to generate an additional simulation. 
This simulation involved $768^3$ particles evolved within a $L=250^3~\mpcht$ box, enabling better visualization of the effects with low computational costs.
With the on-the-fly light-cone code integrated into Gadget-4, we were able to easily obtain full-sky density maps. By stacking the projected density fields with a thickness of $25~\mpch$ at different redshifts, we generated convergence maps that span the entire sky.
We inspect the box-replication effect by showing the lensing convergence maps at $\chi_s = N\times L$, with $N=3, 4, 6$ and $8$.
Twelve maps are presented in Fig.~\ref{fig:visual} to provide visual representations for different $N$ and different directions.
These maps are also generated by selecting a $5 \times 5~\mathrm{deg}^2$ region centered at $(1,0,0)$, $(1,1,0)$ and $(1,1,1)$ direction. 
This visualization clearly shows us the artificial convergence structures depending on the special directions and the number of box repetitions.
Note that $N=3,4$ maps roughly have the same number of replications as the fiducial L1200 case at source redshift $z_s = 1.5, 3.0$, respectively.
These artificial features extend to scale $\theta \sim \pi/\ell \approx 1~\mathrm{deg}$,
explaining the power spectrum excess at $\ell<200$ in Fig.~\ref{fig:masked-ps}.

\subsection{ Convergence Moments }
\label{sec:k234}

\begin{figure*}
    \centering
    \includegraphics[width=0.72\textwidth]{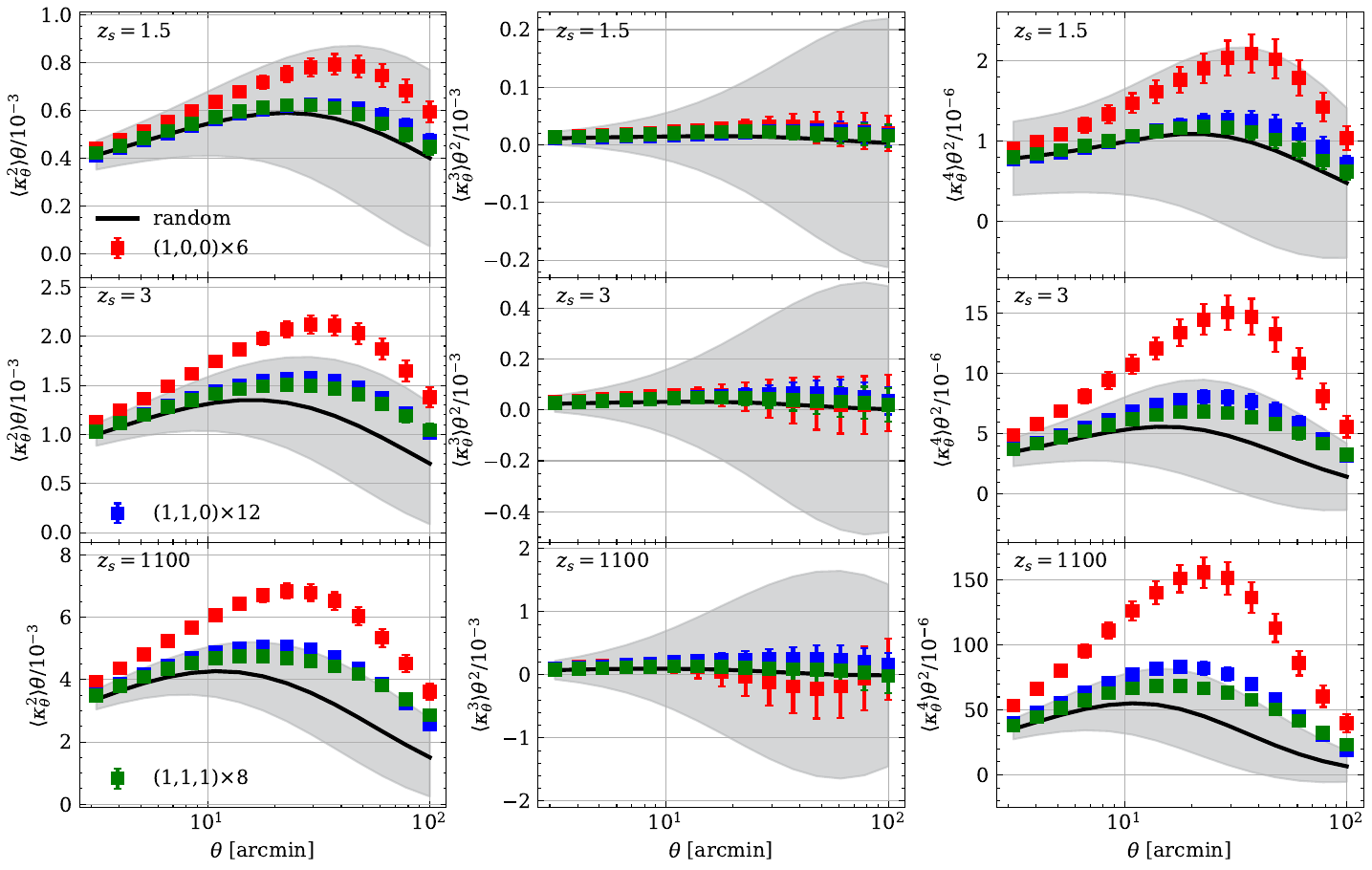}
    \caption{ \textit{Left}: Second-order moments of $25~\mathrm{deg^2}$ convergence fields with different smoothing scales. 
    The results of the source galaxy at $z_s=1.5, 3.0$ are shown in the top and middle panels, respectively. 
    Meanwhile, the CMB lensing fields at $z_s=1100$ are shown in the bottom panel.
    \textit{Middle}: Third-order moments of the smoothed lensing fields.
    \textit{Right}: Fourth-order moments of the smoothed weak lensing and CMB lensing convergence maps.
    }
    \label{fig:k234}
\end{figure*}

After the investigation of the power spectrum, it is natural to inquire whether the box replication effect has a more pronounced influence on higher-order statistics.
Moments of smoothed lensing fields can capture the non-Gaussianity, and definitely will be affected by the non-Gaussian artificial structures appear in Fig.~\ref{fig:visual}.
We smooth the convergence field first and then again apply the mask to pick out a $5\times 5~\mathrm{deg^2}$ region for a given line of sight.
The moments here are less affected by the detailed shape of the mask by definition.
We present the higher-order moments of the smoothed convergence fields for different lensing kernels and line-of-sight directions in Fig.~\ref{fig:k234}.
For the lowest source redshift case, the box replication effect on all statistics is below the statistical uncertainty because of the small times of structure repetition $N \sim 2.6$.
For higher source redshifts, second-order and fourth-order moments are effective in distinguishing $(1,0,0) \times 6$ directions and random viewing angles.
For the other two special orientations, the box replication effect has much lower significance.
This can be attributed to the relatively small repetition times of $N \sim 3.8$.
The kaleidoscopic structures resulting from box tiling appear to be less prominent when $N \lesssim 4$.
All these findings are consistent with the power spectra analysis in the previous section.
Moreover, a comparison can be made between the upper panels and the bottom panels in Fig.~\ref{fig:k234}.
Larger deviations are observed for both the second-order and fourth-order moments of CMB lensing convergence.
We argue that this is primarily due to the monotonically slowly increasing CMB lensing kernels at low redshifts, as shown in Fig.~\ref{fig:Wkappa}.

Surprisingly, when considering third-order moments of smoothed fields alone, it becomes challenging to distinguish between the special directions. 
The box replication effect can create both denser regions and deeper voids in the lensing convergence. 
As a result, this symmetry makes the skewness of $\kappa$ less affected. 

\subsection{ Mask Size }
\label{sec:mask}

In this section, we focus on the mask size in the partial-sky light-cone construction detailed in Sec.~\ref{sec:partial-lc}.
A larger mask will cover more normal regions, and dilute the significance of the box replication effect.
To address this concern, we vary the sky size from $25~\mathrm{deg^2}$ to $400\ \mathrm{deg^2}$ and compute all four statistics for three distinct types of specialized viewing angles, subsequently comparing them to the results obtained from random sampling.
For the partial-sky convergence spectrum, we show the normalized $\hat{C}(\ell) = C(\ell)/ A_{\rm sky}$  at $\ell = 100$ where $A_{\rm sky}$ represents the area of sky coverage.
The smoothing scale of the lensing fields remains fixed at $29~\mathrm{arcmin}$ throughout this analysis, as from Fig.~\ref{fig:k234} at this smoothing scale the deviation is larger than other scales.

\begin{figure*}
    \centering
    \includegraphics[width=0.96\textwidth]{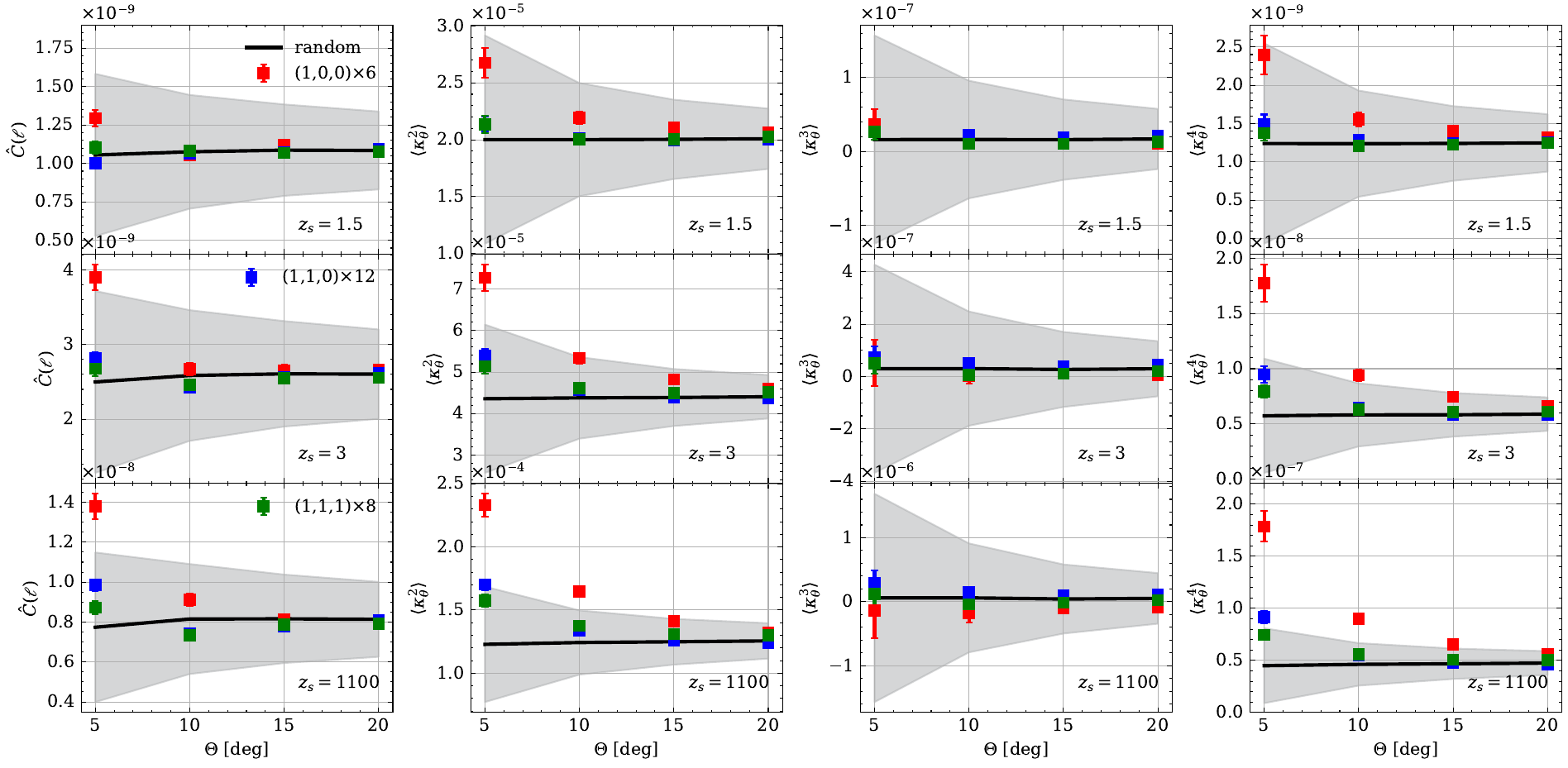}
    \caption{ 
    \textit{First column}: Normalized partial-sky convergence power spectrum at $z_s = 1.5,\ 3.0$ and $1100$.
    The horizontal axis indicates the patch size.
    The measurements of galaxy lensing are presented in the top panel. The results of CMB lensing are shown in the bottom panel.
    \textit{Second column}: Second-order moments of partial-sky convergence fields within different areas.
    \textit{Third column}: Third-order moments of the smoothed lensing fields.
    \textit{Fourth column}: Fourth-order moments of the smoothed weak lensing and CMB lensing convergence maps.}
    \label{fig:k234-masks}
\end{figure*}

All results are shown in Fig.~\ref{fig:k234-masks}.
When the sky coverage is limited, the measurements of moments exhibit noticeable deviations from the expected values.
This deviation is even more pronounced for the data points associated with CMB lensing when compared to low redshift sources.
The trend for different lensing kernels is consistent with the previous conclusion in Sec.~\ref{sec:k234}.
However, as the sky coverage increases to $400~\mathrm{deg^2}$, no significant difference is observed among the various light-cones with different lensing kernels and viewing angles.
This is caused by the decreasing ratio of special areas to total areas when increasing the mask size.
This effect can be statistically insignificant for the Stage-\Rmnum{3} surveys with a few thousand square degrees sky coverage if the simulation volume is sufficiently large ($\gtrsim 0.2 \gpcht $)\footnote{The corresponding side length of the simulation box is $\sim 600~\mpch$, which only reaches four times box replications for galaxy catalog at a $z\sim 1$. 
This should increase according to the redshift distribution for the future survey.
}.
We point out that this dilution in statistics does not mean the systematic features have vanished in special directions.
Therefore, when analyzing local estimators like peaks and minima, additional caution is required.

Fig.~\ref{fig:k234-masks} also implies that the box replication effect diminishes rapidly when the line of sight deviates from certain specific directions.
This suggests that this artificial effect can be avoided as long as we keep away from almost special directions during the light-cone construction.
Therefore, the final problem lies in identifying the directions that give rise to the box replication effect.
We will discuss how to build a robust light-cone by achieving this in Sec.~\ref{sec:losf}.

\subsection{Bayesian Tension Metric}
\label{sec:baysian}

To account for the influence of systematic effects on the theoretical modelling, we adopt a parameter-fitting approach to translate these effects into constraints on cosmological parameters. 
We achieve this by fitting the statistics obtained from the random line-of-sight directions, each with minimized box replication effects, using a univariate model of the form $\boldsymbol{d} = A\boldsymbol{d}_{\rm model}$.
In this model, $\boldsymbol{d}$ represents the measured statistic obtained from each random sample, while $\boldsymbol{d}_{\rm model}$ corresponds to the mean estimation derived from each type of special orientation. The parameter $A$ allows for the quantification of the systematic effects on the observed statistics and their impact on the cosmological parameter constraints. By fitting this model to the data, we can better understand and account for the systematic uncertainties in the theoretical modelling process.

Under the Gaussian distribution assumption, the $A$ parameter is minimized by 
\begin{equation}
\chi^2(A)=\sum_{i, j=1}^{N_{\ell/\theta}}\left(d_i-A d_{\mathrm{model},i}\right)\left[\boldsymbol{C}^{-1}\right]_{i j}\left(d_j-A d_{\mathrm{model}, j}\right).
\label{eq:chisquare}
\end{equation}
We can easily obtain the best-fit parameter by an analytic formula:
\begin{equation}
A=\frac{\boldsymbol{d} \boldsymbol{C}^{-1} \boldsymbol{d}_{\text {model }}}{\boldsymbol{d}_{\text {model }} \boldsymbol{C}^{-1} \boldsymbol{d}_{\text {model }}}.
\label{eq:A}
\end{equation}
Here, $N_{\ell/\theta}$ represents the number of data points for 
the convergence power spectrum and moments.
We use a conservative scale limit at $30 \leq \ell \leq 1000$ with $N_{\ell} = 20$.
It is worth noting that for Stage-\Rmnum{4} surveys, such as those discussed in \citealt{2023MNRAS.tmp.3423Y}, the scale range will extend to $\ell_{\rm max} \sim 3000$, which further reduces the influence of box replication effects due to the inclusion of smaller scales.
For the latter statistics, the smoothing angles are fixed at $[3.2, 100]~\text{arcmin}$ with $N_{\theta}=14$.

Tab.~\ref{tab:bayesian25} depicts the outcomes obtained from the analysis of light-cones with a sky coverage of $5\times 5~\mathrm{deg^2}$. Our findings reveal that the systematic bias resulting from the box replication effects remains smaller than the statistical error, with the exception of the fourth convergence moments of light-cones along the most exceptional direction, denoted as $(1,0,0)$.
Considering the increasing sky coverage of modern galaxy surveys, we have included the results obtained from a $400~\mathrm{deg^2}$ mask in Tab.~\ref{tab:bayesian400}.
Remarkably, when considering the correlation between different scales and statistical errors in our analysis, the influence of the box replication effect on the four lensing statistics can be well controlled.
This moderate impact can be attributed to the dilution of typical regions as the sky coverage expands.
However, these repeated structures persist within these special directions, potentially giving rise to extra angular correlations for other statistics, e.g., peaks, minima, and Minkowski functionals.
We leave this for future investigation.

A recent frequently used convergence field construction scheme is to put the observer at the box corner, and keep the octant sky light-cone (\citep[e.g.,][]{2019ApJS..245...26K,2023MNRAS.524.2556H,2023MNRAS.525.4367H,2024arXiv240513495E}).
Upon examining the two tables, we observe that the standard deviation of $A$, denoted by $\sigma(A)$, exhibits an inverse square root dependence on the sky coverage, satisfying the expectation from a statistical view.
Furthermore, the systematic bias, indicated by $|\langle A \rangle -1|$ decreases with the increasing area if no additional special regions are introduced.
Taking into account the very limited spatial extent of kaleidoscope-like structures and the large fraction of the normal regions, we expect that the systematic bias associated with a single octant light-cone will be significantly reduced compared to the values presented in Tab.~\ref{tab:bayesian400}.

Obviously, the systematic bias induced by the kaleidoscope-like structures depends on the number of box replications $N$, or equivalently the size of the simulation box used to tile the lightcone.
The weak lensing statistics at higher redshift are expected to suffer more from the box replication effect.
However, the measured $\langle A \rangle$ is roughly the same for two source redshifts $z_s = 1.5$ and $3.0$ for given statistics and sky coverage in Tab.~\ref{tab:bayesian25} \&  \ref{tab:bayesian400}.
This coincidence is attributed to two factors, an increased lensing signal at a high source redshift, and an increased systematic bias at a high source redshift.
From Fig.~\ref{fig:visual} we can find that some kaleidoscope feature is composed of a string of radial halo-like structures.
This is a direct consequence of the same halo observed at different redshifts.
They will all contribute to the $\kappa$ moments if the smoothing length is comparable to the scale of these string-like structures.
Naively, the amplitude of bias on $\kappa$ is roughly proportional to an effective $N$ when the lensing kernel is taken into account,
\be
N_\mathrm{eff}=N\frac{\int W(\chi) \dif\chi}{\int W_\mathrm{max}\dif\chi}\ .
\label{eq:effectiveN}
\ee
Here $W_\mathrm{max}$ is the maximum value of the lensing kernel.
For CMB lensing at $z_s=1100$ and weak lensing at $z_s=3$, the simulation boxes employed in the light-cone construction in this work are the same. 
The larger bias that appeared in CMB lensing cases is due to the relatively flat lensing kernel, leading to $N_\mathrm{eff}=0.756N$ compared to the weak lensing result $N_\mathrm{eff}=0.625N$, with $N=3.833$.

\begin{table*}
\begin{tabular}{llccccccccc}
{direction} & {} & \multicolumn{3}{c}{$(1,0,0)$} & \multicolumn{3}{c}{$(1,1,0)$} & \multicolumn{3}{c}{$(1,1,1)$} \\
\cline{1-11}
{statistics} & {} &  $\langle A\rangle$ &  $\sigma(A)$ &  $\frac{|\langle A\rangle - 1|}{\sigma(A)}$ &  $\langle A\rangle$ &  $\sigma(A)$ & $\frac{|\langle A\rangle - 1|}{\sigma(A)}$ &  $\langle A\rangle$ &  $\sigma(A)$ & $\frac{|\langle A\rangle - 1|}{\sigma(A)}$ \\
\cline{1-11}
\multirow{3}{*}{$C_{\kappa\kappa}(\ell)$} &
$z_s = 1.5$  &             0.974928 &      0.115627 &  0.216835 &             1.006055 &      0.118281 &  0.051191 &             0.984959 &      0.115710 &  0.129985 \\
\cline{2-11}
{} & $z_s = 3.0$  &             0.976937 &      0.080271 &  0.287318 &             1.001877 &      0.081253 &  0.023103 &             0.993261 &      0.080433 &  0.083779 \\
\cline{2-11}
{} & $z_s = 1100$ &             0.959066 &      0.060307 &  \bf{0.678757} &             0.992844 &      0.061867 &  0.115666 &             0.994603 &      0.061910 &  0.087167 \\
\cline{1-11}
\multirow{3}{*}{$\langle \kappa^2_{\theta} \rangle$} &
$z_s = 1.5$  &             0.983280 &      0.089468 &  0.186883 &             1.034505 &      0.093977 &  0.367162 &             0.985125 &      0.089406 &  0.166378 \\
\cline{2-11}
{} & $z_s = 3.0$  &             0.979632 &      0.061898 &  0.329055 &             1.020842 &      0.064245 &  0.324409 &             0.989774 &      0.062255 &  0.164258 \\
\cline{2-11}
{} & $z_s = 1100$ &             0.980437 &      0.039573 &  0.494358 &             1.017404 &      0.040917 &  0.425351 &             0.993138 &      0.039938 &  0.171823 \\
\cline{1-11}
\multirow{3}{*}{$\langle \kappa^3_{\theta} \rangle$} &
$z_s = 1.5$  &             0.988661 &      0.255906 &  0.044309 &             1.076210 &      0.277321 &  0.274809 &             0.980875 &      0.253020 &  0.075588 \\
\cline{2-11}
{} & $z_s = 3.0$  &             0.987609 &      0.236556 &  0.052382 &             1.063988 &      0.251966 &  0.253955 &             0.968756 &      0.229322 &  0.136244 \\
\cline{2-11}
{} & $z_s = 1100$ &             0.910621 &      0.244983 &  0.364838 &             1.038345 &      0.271368 &  0.141303 &             0.940633 &      0.246464 &  0.240873 \\
\cline{1-11}
\multirow{3}{*}{$\langle \kappa^4_{\theta} \rangle$} &
$z_s = 1.5$  &             0.668164 &      0.393678 &  \bf{0.842912} &             0.983395 &      0.533746 &  0.031109 &             0.945442 &      0.505685 &  0.107890 \\
\cline{2-11}
{} & $z_s = 3.0$  &             0.568817 &      0.212749 &  \bf{2.026718} &             0.920131 &      0.300942 &  0.265396 &             0.916459 &      0.295385 &  0.282819 \\
\cline{2-11}
{} & $z_s = 1100$ &             0.683477 &      0.126141 &  \bf{2.509281} &             0.950014 &      0.156286 &  0.319838 &             0.924932 &      0.151501 &  0.495497 \\
\cline{1-11}
\end{tabular}
\caption{
The parameter $A$ fitting results for $5\times5~\mathrm{deg^2}$ partial-sky light-cones with three different source redshifts and three kinds of line-of-sight directions.
$\frac{|\langle A \rangle - 1|}{\sigma(A)}$ can be utilized to quantify the systematic bias in the modelling, where $\langle A \rangle$ is the average value of $A$ and $\sigma(A)$ is the standard deviation of $A$.
We use bold font to highlight the bias larger than $0.5 \sigma$.
}
\label{tab:bayesian25}
\end{table*}

\begin{table*}
\begin{tabular}{llccccccccc}
{direction} & {} & \multicolumn{3}{c}{$(1,0,0)$} & \multicolumn{3}{c}{$(1,1,0)$} & \multicolumn{3}{c}{$(1,1,1)$} \\
\cline{1-11}
{statistics} & {} &  $\langle A\rangle$ &  $\sigma(A)$ &  $\frac{|\langle A\rangle - 1|}{\sigma(A)}$ &  $\langle A\rangle$ &  $\sigma(A)$ & $\frac{|\langle A\rangle - 1|}{\sigma(A)}$ &  $\langle A\rangle$ &  $\sigma(A)$ & $\frac{|\langle A\rangle - 1|}{\sigma(A)}$ \\
\cline{1-11}
\multirow{3}{*}{$C_{\kappa\kappa}(\ell)$} &
$z_s = 1.5$  &             1.001167 &      0.032916 &  0.035461 &             1.001257 &      0.032919 &  0.038179 &             1.001737 &      0.032935 &  0.052733 \\
\cline{2-11}
{} & $z_s = 3.0$  &             0.998582 &      0.021475 &  0.066038 &             1.002398 &      0.021557 &  0.111239 &             1.001749 &      0.021543 &  0.081178 \\
\cline{2-11}
{} & $z_s = 1100$ &             0.996026 &      0.015926 &  0.249513 &             1.003449 &      0.016043 &  0.215014 &             1.001949 &      0.016020 &  0.121639 \\
\cline{1-11}
\multirow{3}{*}{$\langle \kappa^2_{\theta} \rangle$} &
$z_s = 1.5$  &             0.996121 &      0.026283 &  0.147605 &             1.011579 &      0.026692 &  0.433797 &             1.002197 &      0.026442 &  0.083106 \\
\cline{2-11}
{} & $z_s = 3.0$  &             0.996252 &      0.017760 &  0.211046 &             1.008455 &      0.017977 &  0.470342 &             1.001049 &      0.017845 &  0.058763 \\
\cline{2-11}
{} & $z_s = 1100$ &             0.995009 &      0.011963 &  0.417200 &             1.008072 &      0.012119 &  \bf{0.666113} &             1.000064 &      0.012023 &  0.005318 \\
\cline{1-11}
\multirow{3}{*}{$\langle \kappa^3_{\theta} \rangle$} &
$z_s = 1.5$  &             0.980255 &      0.065672 &  0.300657 &             1.021229 &      0.068395 &  0.310396 &             1.007512 &      0.067477 &  0.111333 \\
\cline{2-11}
{} & $z_s = 3.0$  &             0.981669 &      0.059531 &  0.307918 &             1.019302 &      0.061784 &  0.312419 &             1.003599 &      0.060840 &  0.059156 \\
\cline{2-11}
{} & $z_s = 1100$ &             0.977213 &      0.063501 &  0.358842 &             1.018759 &      0.066143 &  0.283610 &             0.993324 &      0.064492 &  0.103513 \\
\cline{1-11}
\multirow{3}{*}{$\langle \kappa^4_{\theta} \rangle$} &
$z_s = 1.5$  &             0.965330 &      0.134665 &  0.257456 &             1.014115 &      0.141257 &  0.099921 &             1.019490 &      0.142071 &  0.137184 \\
\cline{2-11}
{} & $z_s = 3.0$  &             0.977436 &      0.084264 &  0.267779 &             1.012450 &      0.087022 &  0.143068 &             1.009962 &      0.086910 &  0.114628 \\
\cline{2-11}
{} & $z_s = 1100$ &             0.989491 &      0.041604 &  0.252600 &             1.014702 &      0.042586 &  0.345230 &             1.002519 &      0.042102 &  0.059836 \\
\cline{1-11}
\end{tabular}
\caption{
Same with Tab.~\ref{tab:bayesian25}, but for a larger sky coverage of $20\times 20~\mathrm{deg^2}$.
}
\label{tab:bayesian400}
\end{table*}

\section{ Line of Sight Finder }
\label{sec:losf}

As discussed in previous sections, it has been established that stacking along the special angles (especially for $(1,0,0)\times6$) leads to artificial structures and obvious systematic errors in higher-order lensing statistics for light-cones with a small coverage.
Fortunately, this effect fades away quickly as indicated in Sec.~\ref{sec:mask}.
This finding allows us to identify alternative special directions that can be explored to minimize the occurrence of excessive box replications.

It is easy to understand that the box replication effect dominates by the times of structure repetitions.
Therefore, we use particles placed on uniform grids to represent small volumes encompassing the entire box.
Then the box is stacked through regular repetition, and the particles are projected onto the lens plane based on a specified light-cone geometry.
Here, we employ the HEALPix scheme to pixelize the entire sky for a given $N_\mathrm{side}$.
In the final step, it is crucial to record the number of appearances for each particle in each pixel.
Since each particle represents a small volume around its position, this allows us to interpret the number of appearances, denoted as $n$, as the number of structure repetitions.
We define the unique index $I = N(n=1)/N(n\geq 1)$ as the quantitative indicator of the box replication effect.
Here $N$ is the number of structures appearing $n$ times in the stacking process.
A unique light-cone is characterized by a value of one for $I$ in a given pixel, indicating the absence of the box replication effect for that particular sky coverage.
To facilitate accessibility, we have made the code publicly available for further exploration and utilization \href{https://github.com/czymh/losf}{\faGithub}.

\begin{figure}
    \centering
    \includegraphics[width=0.45\textwidth]{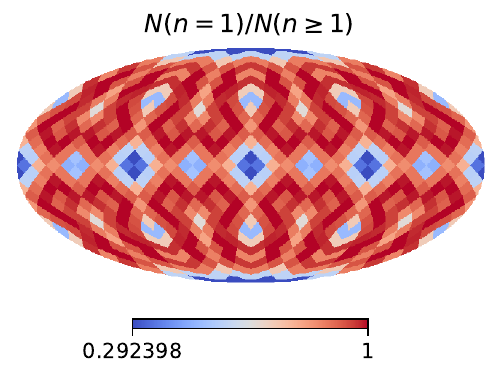}
    \caption{ The full-sky unique index map for $\nside=8$.
    Each pixel represents a pencil-beam light-cone whose sky coverage is about $54~\mathrm{deg^2}$.
    }
    \label{fig:losf}
\end{figure}

By implementing this pipeline, we can conveniently divide the entire sky into 768 pieces uniformly with $\nside = 8$.
The unique indices corresponding to each piece are presented in Fig.~\ref{fig:losf}. 
The light-cone geometry employed here aligns with the convergence map construction of the L1200 simulation.
The underlying test particle number on each dimension is $10$.
We also increase this number to $100$ and obtain the converged result.
In this case, there are $128$ unique light-cones, and the total sky coverage amounts to $6875~\mathrm{deg^2}$.
Notably, the directions characterized by high repetition times, indicated by the blue region in the figure, are primarily concentrated around three distinct types of special directions.
This is also consistent with our previous findings, further supporting our earlier analysis and conclusions.


\section{Conclusions}
\label{sec:conclusion}

Within the scope of this study, we have examined the influence of the box replication effect on convergence power spectra and high-order moments of lensing fields.
The origin of this effect can be traced back to the finite simulated volume, thereby impacting both post-processing and on-the-fly light-cone constructions.

To satisfy the requirement of Stage-\Rmnum{4} surveys, we used the high-resolution N-body simulation evolving $3072^3$ particles with large box sizes $1.2\ \gpch$ to construct light-cones along different directions by tiling the simulation boxes.
The measurements of convergence power spectra and higher-order moments of the light-cones along three kinds of special orientations are compared with the results of random line-of-sight directions in Sec.~\ref{sec:results}.
For small patch size $25~\deg^2$, noticeable deviations can be observed in the power spectrum at $\ell<200$, as well as in second and fourth-order moments for $z_s=3$ and $1100$.
However, the impact of the box replication effect remains confined within the bounds of statistic error for these statistics with $z_s = 1.5$ as shown in Fig.~\ref{fig:masked-ps} and Fig.~\ref{fig:k234}.
The visual inspection reveals that the tessellation geometry employed in weak gravitational lensing simulations can introduce kaleidoscopic structures, particularly in directions where multiple copies of the simulation box are present at high source redshifts. 

To assess the systematic bias in weak lensing statistics, a univariate model is utilized to fit estimations obtained from random samples to measurements derived from light-cones oriented towards specific viewing angles.
Significant bias arising from the aforementioned repeated structures is observed primarily in the fourth-order moments of smoothed CMB lensing convergence fields. 
However, this bias diminishes below the level of statistical errors as the size of the mask is increased from $25~\mathrm{deg^2}$ to $400~\mathrm{deg^2}$.
Consequently, it is concluded that the impact of the box replication effect on the four lensing statistics is statistically insignificant for Stage-\Rmnum{3} and Stage-\Rmnum{4} surveys when employing current dark matter simulations featuring a box size of $L \gtrsim 600 \mpch$.
Nonetheless, this effect becomes significant because of the extended box replication time $N$ for the simulation with $L \lesssim 250 \mpch$, a prevalent choice for hydrodynamic simulations.
Given the presence of repeated structures along special directions, caution is advised when evaluating high-order statistics derived from light-cones with limited sky coverage. 

In such instances, these artificial features rapidly vanish when the viewing angles deviate from specific directions, as indicated in Figure \ref{fig:k234-masks}. 
Thus, unbiased construction of light-cones can be achieved by avoiding these particular lines of sight.
We developed a code to find the unique light-cones to avoid structure repetition spanning the whole sky and make it publicly available. 
In summary, the box replication effect is trivial for generating a mock of the current galaxy survey (effectively $z_s \sim 1.5$) through cosmological simulation with a large box ($L \gtrsim 600~\mpch$) \citep[e.g.,][]{2021JCAP...02..047S,2023JCAP...02..050K,2023MNRAS.524.5591F,2024MNRAS.529.2309B, 2024MNRAS.530.5030O}.

For the purpose of precision cosmological constraints, the box replication effect on the covariance of different lensing statistics should also be quantified well.
The generation of multiple realizations of lensing convergence maps is indeed necessary for estimating covariance matrices precisely. 
Therefore, we defer this task to future investigation.
Another possible extension of this work is to investigate the box replication effect on other statistics, especially for some morphological tools, i.e., Minkowski functionals.
The incorporation of novel statistical methodologies should facilitate a more comprehensive modelling of these anomalous artificial features.

\section*{Acknowledgements}

We thank Pengjie Zhang for the useful discussions.
This work is supported by the National Key R\&D Program of China (2023YFA1607800, 2023YFA1607802), 
the National Science Foundation of China (grant number 12273020), 
the China Manned Space Project with numbers CMS-CSST-2021-A03, 
the ‘111’ Project of the Ministry of Education under grant number B20019, 
and the sponsorship from the Yangyang Development Fund. 
This work made use of the Gravity Supercomputer at the Department of Astronomy, Shanghai Jiao Tong University.

\section*{Data Availability}
All data included in this study are available upon reasonable request by contacting the corresponding author.



\bibliographystyle{mnras}
\bibliography{ref} 

\begin{thebibliography}{}
\makeatletter
\relax
\def\mn@urlcharsother{\let\do\@makeother \do\$\do\&\do\#\do\^\do\_\do\%\do\~}
\def\mn@doi{\begingroup\mn@urlcharsother \@ifnextchar [ {\mn@doi@}
  {\mn@doi@[]}}
\def\mn@doi@[#1]#2{\def\@tempa{#1}\ifx\@tempa\@empty \href
  {http://dx.doi.org/#2} {doi:#2}\else \href {http://dx.doi.org/#2} {#1}\fi
  \endgroup}
\def\mn@eprint#1#2{\mn@eprint@#1:#2::\@nil}
\def\mn@eprint@arXiv#1{\href {http://arxiv.org/abs/#1} {{\tt arXiv:#1}}}
\def\mn@eprint@dblp#1{\href {http://dblp.uni-trier.de/rec/bibtex/#1.xml}
  {dblp:#1}}
\def\mn@eprint@#1:#2:#3:#4\@nil{\def\@tempa {#1}\def\@tempb {#2}\def\@tempc
  {#3}\ifx \@tempc \@empty \let \@tempc \@tempb \let \@tempb \@tempa \fi \ifx
  \@tempb \@empty \def\@tempb {arXiv}\fi \@ifundefined
  {mn@eprint@\@tempb}{\@tempb:\@tempc}{\expandafter \expandafter \csname
  mn@eprint@\@tempb\endcsname \expandafter{\@tempc}}}

\bibitem[\protect\citeauthoryear{{Angulo} \& {Hahn}}{{Angulo} \&
  {Hahn}}{2022}]{2022LRCA....8....1A}
{Angulo} R.~E.,  {Hahn} O.,  2022, \mn@doi [Living Reviews in Computational
  Astrophysics] {10.1007/s41115-021-00013-z}, \href
  {https://ui.adsabs.harvard.edu/abs/2022LRCA....8....1A} {8, 1}

\bibitem[\protect\citeauthoryear{{Bartelmann} \& {Schneider}}{{Bartelmann} \&
  {Schneider}}{2001}]{2001PhR...340..291B}
{Bartelmann} M.,  {Schneider} P.,  2001, \mn@doi [\physrep]
  {10.1016/S0370-1573(00)00082-X}, \href
  {https://ui.adsabs.harvard.edu/abs/2001PhR...340..291B} {340, 291}

\bibitem[\protect\citeauthoryear{{Bernyk} et~al.,}{{Bernyk}
  et~al.}{2016}]{2016ApJS..223....9B}
{Bernyk} M.,  et~al., 2016, \mn@doi [\apjs] {10.3847/0067-0049/223/1/9}, \href
  {https://ui.adsabs.harvard.edu/abs/2016ApJS..223....9B} {223, 9}

\bibitem[\protect\citeauthoryear{{Blaizot}, {Wadadekar}, {Guiderdoni},
  {Colombi}, {Bertin}, {Bouchet}, {Devriendt}  \& {Hatton}}{{Blaizot}
  et~al.}{2005}]{2005MNRAS.360..159B}
{Blaizot} J.,  {Wadadekar} Y.,  {Guiderdoni} B.,  {Colombi} S.~T.,  {Bertin}
  E.,  {Bouchet} F.~R.,  {Devriendt} J. E.~G.,   {Hatton} S.,  2005, \mn@doi
  [\mnras] {10.1111/j.1365-2966.2005.09019.x}, \href
  {https://ui.adsabs.harvard.edu/abs/2005MNRAS.360..159B} {360, 159}

\bibitem[\protect\citeauthoryear{{Broxterman} et~al.,}{{Broxterman}
  et~al.}{2024}]{2024MNRAS.529.2309B}
{Broxterman} J.~C.,  et~al., 2024, \mn@doi [\mnras] {10.1093/mnras/stae698},
  \href {https://ui.adsabs.harvard.edu/abs/2024MNRAS.529.2309B} {529, 2309}

\bibitem[\protect\citeauthoryear{{Carlson} \& {White}}{{Carlson} \&
  {White}}{2010}]{2010ApJS..190..311C}
{Carlson} J.,  {White} M.,  2010, \mn@doi [\apjs]
  {10.1088/0067-0049/190/2/311}, \href
  {https://ui.adsabs.harvard.edu/abs/2010ApJS..190..311C} {190, 311}

\bibitem[\protect\citeauthoryear{{Chen}, {Yu}, {Liu}  \& {Fan}}{{Chen}
  et~al.}{2020}]{2020ApJ...897...14C}
{Chen} Z.,  {Yu} Y.,  {Liu} X.,   {Fan} Z.,  2020, \mn@doi [\apj]
  {10.3847/1538-4357/ab980f}, \href
  {https://ui.adsabs.harvard.edu/abs/2020ApJ...897...14C} {897, 14}

\bibitem[\protect\citeauthoryear{{Cooray} \& {Hu}}{{Cooray} \&
  {Hu}}{2002}]{Cooray:2002aa}
{Cooray} A.,  {Hu} W.,  2002, \mn@doi [\apj] {10.1086/340892}, \href
  {https://ui.adsabs.harvard.edu/abs/2002ApJ...574...19C} {574, 19}

\bibitem[\protect\citeauthoryear{{Euclid Collaboration} et~al.,}{{Euclid
  Collaboration} et~al.}{2024}]{2024arXiv240513495E}
{Euclid Collaboration} et~al., 2024, \mn@doi [arXiv e-prints]
  {10.48550/arXiv.2405.13495}, \href
  {https://ui.adsabs.harvard.edu/abs/2024arXiv240513495E} {p. arXiv:2405.13495}

\bibitem[\protect\citeauthoryear{{Evrard} et~al.,}{{Evrard}
  et~al.}{2002}]{2002ApJ...573....7E}
{Evrard} A.~E.,  et~al., 2002, \mn@doi [\apj] {10.1086/340551}, \href
  {https://ui.adsabs.harvard.edu/abs/2002ApJ...573....7E} {573, 7}

\bibitem[\protect\citeauthoryear{{Ferlito} et~al.,}{{Ferlito}
  et~al.}{2023}]{2023MNRAS.524.5591F}
{Ferlito} F.,  et~al., 2023, \mn@doi [\mnras] {10.1093/mnras/stad2205}, \href
  {https://ui.adsabs.harvard.edu/abs/2023MNRAS.524.5591F} {524, 5591}

\bibitem[\protect\citeauthoryear{{Fosalba}, {Gazta{\~n}aga}, {Castander}  \&
  {Manera}}{{Fosalba} et~al.}{2008}]{2008MNRAS.391..435F}
{Fosalba} P.,  {Gazta{\~n}aga} E.,  {Castander} F.~J.,   {Manera} M.,  2008,
  \mn@doi [\mnras] {10.1111/j.1365-2966.2008.13910.x}, \href
  {https://ui.adsabs.harvard.edu/abs/2008MNRAS.391..435F} {391, 435}

\bibitem[\protect\citeauthoryear{{Gao} et~al.,}{{Gao}
  et~al.}{2023}]{2023ApJ...954..207G}
{Gao} H.,  et~al., 2023, \mn@doi [\apj] {10.3847/1538-4357/ace90a}, \href
  {https://ui.adsabs.harvard.edu/abs/2023ApJ...954..207G} {954, 207}

\bibitem[\protect\citeauthoryear{{Gatti} et~al.,}{{Gatti}
  et~al.}{2020}]{2020MNRAS.498.4060G}
{Gatti} M.,  et~al., 2020, \mn@doi [\mnras] {10.1093/mnras/staa2680}, \href
  {https://ui.adsabs.harvard.edu/abs/2020MNRAS.498.4060G} {498, 4060}

\bibitem[\protect\citeauthoryear{{Gatti} et~al.,}{{Gatti}
  et~al.}{2022}]{2022PhRvD.106h3509G}
{Gatti} M.,  et~al., 2022, \mn@doi [\prd] {10.1103/PhysRevD.106.083509}, \href
  {https://ui.adsabs.harvard.edu/abs/2022PhRvD.106h3509G} {106, 083509}

\bibitem[\protect\citeauthoryear{{Giocoli}, {Baldi}  \& {Moscardini}}{{Giocoli}
  et~al.}{2018}]{2018MNRAS.481.2813G}
{Giocoli} C.,  {Baldi} M.,   {Moscardini} L.,  2018, \mn@doi [\mnras]
  {10.1093/mnras/sty2465}, \href
  {https://ui.adsabs.harvard.edu/abs/2018MNRAS.481.2813G} {481, 2813}

\bibitem[\protect\citeauthoryear{{G{\'o}rski}, {Hivon}, {Banday}, {Wandelt},
  {Hansen}, {Reinecke}  \& {Bartelmann}}{{G{\'o}rski}
  et~al.}{2005}]{2005ApJ...622..759G}
{G{\'o}rski} K.~M.,  {Hivon} E.,  {Banday} A.~J.,  {Wandelt} B.~D.,  {Hansen}
  F.~K.,  {Reinecke} M.,   {Bartelmann} M.,  2005, \mn@doi [\apj]
  {10.1086/427976}, \href
  {https://ui.adsabs.harvard.edu/abs/2005ApJ...622..759G} {622, 759}

\bibitem[\protect\citeauthoryear{{Hadzhiyska} et~al.,}{{Hadzhiyska}
  et~al.}{2023}]{2023MNRAS.525.4367H}
{Hadzhiyska} B.,  et~al., 2023, \mn@doi [\mnras] {10.1093/mnras/stad2563},
  \href {https://ui.adsabs.harvard.edu/abs/2023MNRAS.525.4367H} {525, 4367}

\bibitem[\protect\citeauthoryear{{Hamana} et~al.,}{{Hamana}
  et~al.}{2020}]{2020PASJ...72...16H}
{Hamana} T.,  et~al., 2020, \mn@doi [\pasj] {10.1093/pasj/psz138}, \href
  {https://ui.adsabs.harvard.edu/abs/2020PASJ...72...16H} {72, 16}

\bibitem[\protect\citeauthoryear{{Harnois-D{\'e}raps} \& {van
  Waerbeke}}{{Harnois-D{\'e}raps} \& {van
  Waerbeke}}{2015}]{Harnois-Deraps:2015aa}
{Harnois-D{\'e}raps} J.,  {van Waerbeke} L.,  2015, \mn@doi [\mnras]
  {10.1093/mnras/stv794}, \href
  {https://ui.adsabs.harvard.edu/abs/2015MNRAS.450.2857H} {450, 2857}

\bibitem[\protect\citeauthoryear{{Hern{\'a}ndez-Aguayo}
  et~al.,}{{Hern{\'a}ndez-Aguayo} et~al.}{2023}]{2023MNRAS.524.2556H}
{Hern{\'a}ndez-Aguayo} C.,  et~al., 2023, \mn@doi [\mnras]
  {10.1093/mnras/stad1657}, \href
  {https://ui.adsabs.harvard.edu/abs/2023MNRAS.524.2556H} {524, 2556}

\bibitem[\protect\citeauthoryear{{Hilbert}, {Hartlap}, {White}  \&
  {Schneider}}{{Hilbert} et~al.}{2009}]{2009A&A...499...31H}
{Hilbert} S.,  {Hartlap} J.,  {White} S.~D.~M.,   {Schneider} P.,  2009,
  \mn@doi [\aap] {10.1051/0004-6361/200811054}, \href
  {https://ui.adsabs.harvard.edu/abs/2009A&A...499...31H} {499, 31}

\bibitem[\protect\citeauthoryear{{Hilbert} et~al.,}{{Hilbert}
  et~al.}{2020}]{2020MNRAS.493..305H}
{Hilbert} S.,  et~al., 2020, \mn@doi [\mnras] {10.1093/mnras/staa281}, \href
  {https://ui.adsabs.harvard.edu/abs/2020MNRAS.493..305H} {493, 305}

\bibitem[\protect\citeauthoryear{{Hildebrandt} et~al.,}{{Hildebrandt}
  et~al.}{2017}]{2017MNRAS.465.1454H}
{Hildebrandt} H.,  et~al., 2017, \mn@doi [\mnras] {10.1093/mnras/stw2805},
  \href {https://ui.adsabs.harvard.edu/abs/2017MNRAS.465.1454H} {465, 1454}

\bibitem[\protect\citeauthoryear{{Hinshaw} et~al.,}{{Hinshaw}
  et~al.}{2013}]{2013ApJS..208...19H}
{Hinshaw} G.,  et~al., 2013, \mn@doi [\apjs] {10.1088/0067-0049/208/2/19},
  \href {https://ui.adsabs.harvard.edu/abs/2013ApJS..208...19H} {208, 19}

\bibitem[\protect\citeauthoryear{{Izquierdo-Villalba}
  et~al.,}{{Izquierdo-Villalba} et~al.}{2019}]{2019A&A...631A..82I}
{Izquierdo-Villalba} D.,  et~al., 2019, \mn@doi [\aap]
  {10.1051/0004-6361/201936232}, \href
  {https://ui.adsabs.harvard.edu/abs/2019A&A...631A..82I} {631, A82}

\bibitem[\protect\citeauthoryear{{Jain}, {Seljak}  \& {White}}{{Jain}
  et~al.}{2000}]{2000ApJ...530..547J}
{Jain} B.,  {Seljak} U.,   {White} S.,  2000, \mn@doi [\apj] {10.1086/308384},
  \href {https://ui.adsabs.harvard.edu/abs/2000ApJ...530..547J} {530, 547}

\bibitem[\protect\citeauthoryear{{Jing}}{{Jing}}{2019}]{Jing:2019uw}
{Jing} Y.,  2019, \mn@doi [Science China Physics, Mechanics, and Astronomy]
  {10.1007/s11433-018-9286-x}, \href
  {https://ui.adsabs.harvard.edu/abs/2019SCPMA..6219511J} {62, 19511}

\bibitem[\protect\citeauthoryear{{Kacprzak}, {Fluri}, {Schneider}, {Refregier}
  \& {Stadel}}{{Kacprzak} et~al.}{2023}]{2023JCAP...02..050K}
{Kacprzak} T.,  {Fluri} J.,  {Schneider} A.,  {Refregier} A.,   {Stadel} J.,
  2023, \mn@doi [\jcap] {10.1088/1475-7516/2023/02/050}, \href
  {https://ui.adsabs.harvard.edu/abs/2023JCAP...02..050K} {2023, 050}

\bibitem[\protect\citeauthoryear{{Kilbinger}}{{Kilbinger}}{2015}]{2015RPPh...78h6901K}
{Kilbinger} M.,  2015, \mn@doi [Reports on Progress in Physics]
  {10.1088/0034-4885/78/8/086901}, \href
  {https://ui.adsabs.harvard.edu/abs/2015RPPh...78h6901K} {78, 086901}

\bibitem[\protect\citeauthoryear{{Kitzbichler} \& {White}}{{Kitzbichler} \&
  {White}}{2007}]{2007MNRAS.376....2K}
{Kitzbichler} M.~G.,  {White} S.~D.~M.,  2007, \mn@doi [\mnras]
  {10.1111/j.1365-2966.2007.11458.x}, \href
  {https://ui.adsabs.harvard.edu/abs/2007MNRAS.376....2K} {376, 2}

\bibitem[\protect\citeauthoryear{{Klypin} \& {Prada}}{{Klypin} \&
  {Prada}}{2019}]{2019MNRAS.489.1684K}
{Klypin} A.,  {Prada} F.,  2019, \mn@doi [\mnras] {10.1093/mnras/stz2194},
  \href {https://ui.adsabs.harvard.edu/abs/2019MNRAS.489.1684K} {489, 1684}

\bibitem[\protect\citeauthoryear{{Komatsu} et~al.,}{{Komatsu}
  et~al.}{2011}]{2011ApJS..192...18K}
{Komatsu} E.,  et~al., 2011, \mn@doi [\apjs] {10.1088/0067-0049/192/2/18},
  \href {https://ui.adsabs.harvard.edu/abs/2011ApJS..192...18K} {192, 18}

\bibitem[\protect\citeauthoryear{{Korytov} et~al.,}{{Korytov}
  et~al.}{2019}]{2019ApJS..245...26K}
{Korytov} D.,  et~al., 2019, \mn@doi [\apjs] {10.3847/1538-4365/ab510c}, \href
  {https://ui.adsabs.harvard.edu/abs/2019ApJS..245...26K} {245, 26}

\bibitem[\protect\citeauthoryear{{Mandelbaum}}{{Mandelbaum}}{2018}]{2018ARA&A..56..393M}
{Mandelbaum} R.,  2018, \mn@doi [\araa] {10.1146/annurev-astro-081817-051928},
  \href {https://ui.adsabs.harvard.edu/abs/2018ARA&A..56..393M} {56, 393}

\bibitem[\protect\citeauthoryear{{Mead}, {Peacock}, {Heymans}, {Joudaki}  \&
  {Heavens}}{{Mead} et~al.}{2015}]{2015MNRAS.454.1958M}
{Mead} A.~J.,  {Peacock} J.~A.,  {Heymans} C.,  {Joudaki} S.,   {Heavens}
  A.~F.,  2015, \mn@doi [\mnras] {10.1093/mnras/stv2036}, \href
  {https://ui.adsabs.harvard.edu/abs/2015MNRAS.454.1958M} {454, 1958}

\bibitem[\protect\citeauthoryear{{Merson} et~al.,}{{Merson}
  et~al.}{2013}]{2013MNRAS.429..556M}
{Merson} A.~I.,  et~al., 2013, \mn@doi [\mnras] {10.1093/mnras/sts355}, \href
  {https://ui.adsabs.harvard.edu/abs/2013MNRAS.429..556M} {429, 556}

\bibitem[\protect\citeauthoryear{{Munshi} \& {Valageas}}{{Munshi} \&
  {Valageas}}{2005}]{2005astro.ph.10266M}
{Munshi} D.,  {Valageas} P.,  2005, arXiv e-prints, \href
  {https://ui.adsabs.harvard.edu/abs/2005astro.ph.10266M} {pp
  astro--ph/0510266}

\bibitem[\protect\citeauthoryear{{Omori}}{{Omori}}{2024}]{2024MNRAS.530.5030O}
{Omori} Y.,  2024, \mn@doi [\mnras] {10.1093/mnras/stae1031}, \href
  {https://ui.adsabs.harvard.edu/abs/2024MNRAS.530.5030O} {530, 5030}

\bibitem[\protect\citeauthoryear{{Petri}, {Liu}, {Haiman}, {May}, {Hui}  \&
  {Kratochvil}}{{Petri} et~al.}{2015}]{2015PhRvD..91j3511P}
{Petri} A.,  {Liu} J.,  {Haiman} Z.,  {May} M.,  {Hui} L.,   {Kratochvil}
  J.~M.,  2015, \mn@doi [\prd] {10.1103/PhysRevD.91.103511}, \href
  {https://ui.adsabs.harvard.edu/abs/2015PhRvD..91j3511P} {91, 103511}

\bibitem[\protect\citeauthoryear{{Potter}, {Stadel}  \& {Teyssier}}{{Potter}
  et~al.}{2017}]{2017ComAC...4....2P}
{Potter} D.,  {Stadel} J.,   {Teyssier} R.,  2017, \mn@doi [Computational
  Astrophysics and Cosmology] {10.1186/s40668-017-0021-1}, \href
  {https://ui.adsabs.harvard.edu/abs/2017ComAC...4....2P} {4, 2}

\bibitem[\protect\citeauthoryear{{Rodr{\'\i}guez-Torres}
  et~al.,}{{Rodr{\'\i}guez-Torres} et~al.}{2016}]{2016MNRAS.460.1173R}
{Rodr{\'\i}guez-Torres} S.~A.,  et~al., 2016, \mn@doi [\mnras]
  {10.1093/mnras/stw1014}, \href
  {https://ui.adsabs.harvard.edu/abs/2016MNRAS.460.1173R} {460, 1173}

\bibitem[\protect\citeauthoryear{{Schneider} et~al.,}{{Schneider}
  et~al.}{2016}]{2016JCAP...04..047S}
{Schneider} A.,  et~al., 2016, \mn@doi [\jcap] {10.1088/1475-7516/2016/04/047},
  \href {https://ui.adsabs.harvard.edu/abs/2016JCAP...04..047S} {2016, 047}

\bibitem[\protect\citeauthoryear{{Sgier}, {R{\'e}fr{\'e}gier}, {Amara}  \&
  {Nicola}}{{Sgier} et~al.}{2019}]{2019JCAP...01..044S}
{Sgier} R.~J.,  {R{\'e}fr{\'e}gier} A.,  {Amara} A.,   {Nicola} A.,  2019,
  \mn@doi [\jcap] {10.1088/1475-7516/2019/01/044}, \href
  {https://ui.adsabs.harvard.edu/abs/2019JCAP...01..044S} {2019, 044}

\bibitem[\protect\citeauthoryear{{Sgier}, {Fluri}, {Herbel},
  {R{\'e}fr{\'e}gier}, {Amara}, {Kacprzak}  \& {Nicola}}{{Sgier}
  et~al.}{2021}]{2021JCAP...02..047S}
{Sgier} R.,  {Fluri} J.,  {Herbel} J.,  {R{\'e}fr{\'e}gier} A.,  {Amara} A.,
  {Kacprzak} T.,   {Nicola} A.,  2021, \mn@doi [\jcap]
  {10.1088/1475-7516/2021/02/047}, \href
  {https://ui.adsabs.harvard.edu/abs/2021JCAP...02..047S} {2021, 047}

\bibitem[\protect\citeauthoryear{{Smith}, {Cole}, {Baugh}, {Zheng}, {Angulo},
  {Norberg}  \& {Zehavi}}{{Smith} et~al.}{2017}]{2017MNRAS.470.4646S}
{Smith} A.,  {Cole} S.,  {Baugh} C.,  {Zheng} Z.,  {Angulo} R.,  {Norberg} P.,
   {Zehavi} I.,  2017, \mn@doi [\mnras] {10.1093/mnras/stx1432}, \href
  {https://ui.adsabs.harvard.edu/abs/2017MNRAS.470.4646S} {470, 4646}

\bibitem[\protect\citeauthoryear{Sobol'}{Sobol'}{1967}]{SOBOL196786}
Sobol' I.,  1967, \mn@doi [USSR Computational Mathematics and Mathematical
  Physics] {https://doi.org/10.1016/0041-5553(67)90144-9}, 7, 86

\bibitem[\protect\citeauthoryear{{Springel}, {Pakmor}, {Zier}  \&
  {Reinecke}}{{Springel} et~al.}{2021}]{2021MNRAS.506.2871S}
{Springel} V.,  {Pakmor} R.,  {Zier} O.,   {Reinecke} M.,  2021, \mn@doi
  [\mnras] {10.1093/mnras/stab1855}, \href
  {https://ui.adsabs.harvard.edu/abs/2021MNRAS.506.2871S} {506, 2871}

\bibitem[\protect\citeauthoryear{{Takahashi}, {Hamana}, {Shirasaki},
  {Namikawa}, {Nishimichi}, {Osato}  \& {Shiroyama}}{{Takahashi}
  et~al.}{2017}]{2017ApJ...850...24T}
{Takahashi} R.,  {Hamana} T.,  {Shirasaki} M.,  {Namikawa} T.,  {Nishimichi}
  T.,  {Osato} K.,   {Shiroyama} K.,  2017, \mn@doi [\apj]
  {10.3847/1538-4357/aa943d}, \href
  {https://ui.adsabs.harvard.edu/abs/2017ApJ...850...24T} {850, 24}

\bibitem[\protect\citeauthoryear{{Troxel} et~al.,}{{Troxel}
  et~al.}{2018}]{2018PhRvD..98d3528T}
{Troxel} M.~A.,  et~al., 2018, \mn@doi [\prd] {10.1103/PhysRevD.98.043528},
  \href {https://ui.adsabs.harvard.edu/abs/2018PhRvD..98d3528T} {98, 043528}

\bibitem[\protect\citeauthoryear{{Upadhye} et~al.,}{{Upadhye}
  et~al.}{2024}]{2024MNRAS.529.1862U}
{Upadhye} A.,  et~al., 2024, \mn@doi [\mnras] {10.1093/mnras/stae663}, \href
  {https://ui.adsabs.harvard.edu/abs/2024MNRAS.529.1862U} {529, 1862}

\bibitem[\protect\citeauthoryear{{Wei} et~al.,}{{Wei}
  et~al.}{2018}]{2018ApJ...853...25W}
{Wei} C.,  et~al., 2018, \mn@doi [\apj] {10.3847/1538-4357/aaa40d}, \href
  {https://ui.adsabs.harvard.edu/abs/2018ApJ...853...25W} {853, 25}

\bibitem[\protect\citeauthoryear{{Yao} et~al.,}{{Yao}
  et~al.}{2023}]{2023MNRAS.tmp.3423Y}
{Yao} J.,  et~al., 2023, \mn@doi [\mnras] {10.1093/mnras/stad3563}, \href
  {https://ui.adsabs.harvard.edu/abs/2023MNRAS.tmp.3423Y} {}

\bibitem[\protect\citeauthoryear{{Zorrilla Matilla}, {Waterval}  \&
  {Haiman}}{{Zorrilla Matilla} et~al.}{2020}]{2020AJ....159..284Z}
{Zorrilla Matilla} J.~M.,  {Waterval} S.,   {Haiman} Z.,  2020, \mn@doi [\aj]
  {10.3847/1538-3881/ab8f8c}, \href
  {https://ui.adsabs.harvard.edu/abs/2020AJ....159..284Z} {159, 284}

\makeatother
\end{thebibliography}




\appendix

\section{Randomisation Procedure}
\label{app:random}

In this appendix, we examine the impact of three different rotation strategies on the convergence power spectrum.
\begin{itemize}
    \item No Rotation: No rotation is applied to each mass shell which is our fiducial choice in the main text.
    \item Box Rotation: Shells are randomly rotated in bundles of a box length $1200~\mpch$.
    \item Shell Rotation: Each mass shell undergoes a random rotation to mitigate repeated structures in the light-cone construction.
\end{itemize}
The mass shells pixelated by HEALPix are rotated in the spherical harmonic space rather than rotating the particles directly to save computational cost.
To reduce cosmic variance, we generate 25 realizations for each of the three methods.

\begin{figure*}
    \centering
    \includegraphics[width=0.96\textwidth]{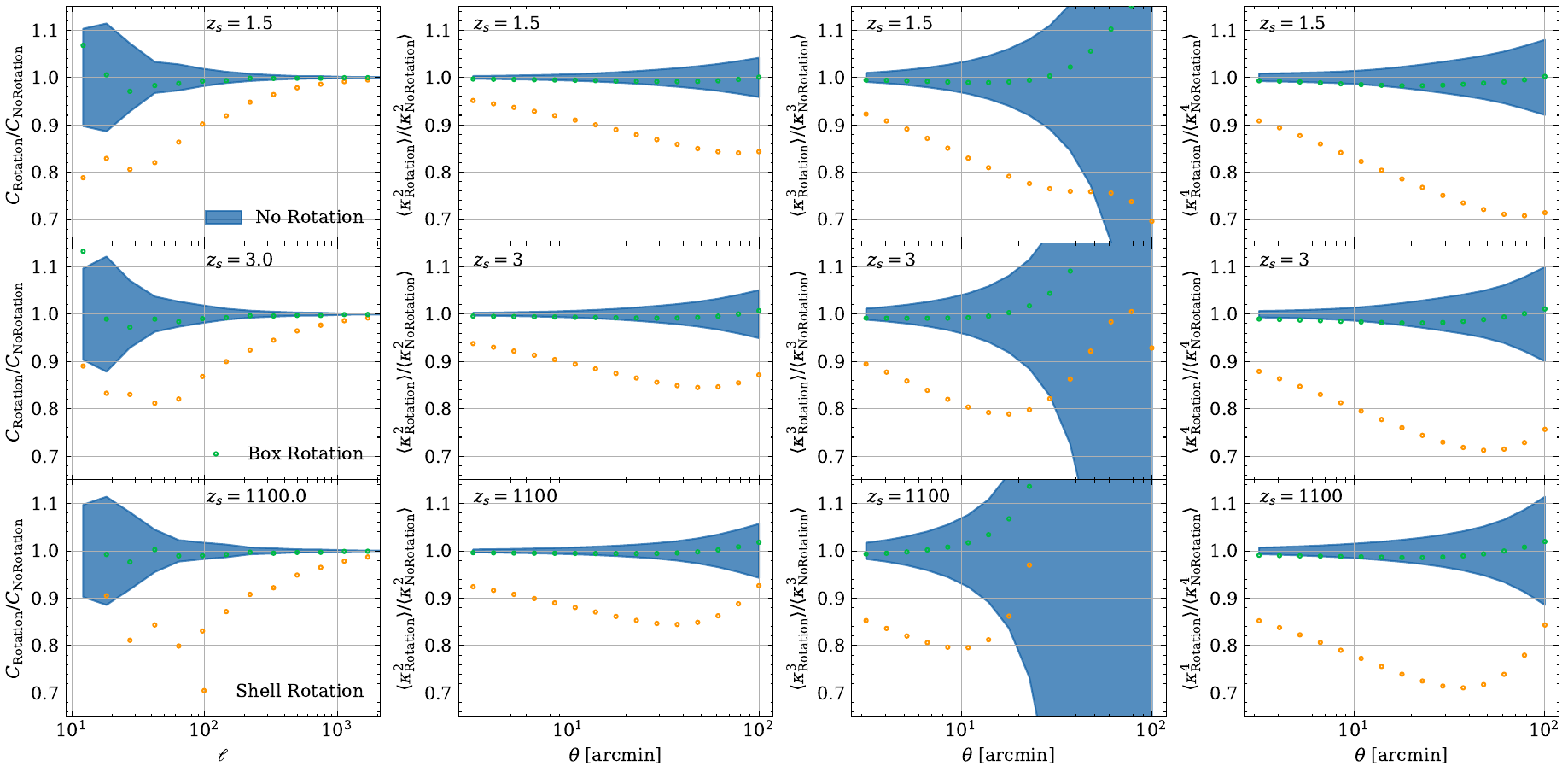}
    \caption{The impact of three rotation strategies on four statistics for source galaxies at $z_s = 1.5, 3.0, 1100$.
    The first to fourth columns present the results of the power spectrum, second-order moments, third-order moments, and fourth-order moments, respectively.
    The mean measurement of 25 non-rotation light-cones is shown by the blue shadow, green circles for the rotated shells for every box length, and yellow ones for each rotated shell. 
    The error bars represent the fluctuation of all realizations.
    }
    \label{fig:rotated-full-sky-cl}
\end{figure*}

Fig.~\ref{fig:rotated-full-sky-cl} illustrates the ratios of full-sky convergence power spectrum and moments for these strategies relative to no-rotation results at three source redshifts.
In the first to fourth columns, the results of the power spectrum, second-order moments, third-order moments, and fourth-order moments are shown in sequence. 
Correspondingly, from the first to the third rows, the data for the source redshifts at 1.5, 3.0, and 1100 are exhibited in turn.
The mean measurement and statistical error of no rotation results is shown by the blue shadow. 
In contrast, the rotation of every box length and the rotation of each individual shell are indicated by green and yellow circles, respectively.

For the power spectrum, the approximately $2\%$ suppression at $\ell \lesssim 100$ introduced by the second rotation strategy is negligible compared to cosmic variance. 
However, for second and fourth-order moments, we observe that this strategy induces $1\sigma$ level deviations for $\theta<10~\arcmin$.
Rotating each mass shell in the light-cone construction has a considerable impact on large scales for all statistics.
The power spectrum result here is consistent with \citealt{2024MNRAS.529.1862U}.
The more times of rotations in this method result in a greater loss of correlation between different mass shells.
Given that the rotation changes the amplitude of the high-order moments comparable to the statistical uncertainty at small scales, we do not utilize the rotation method to set the reference.

\section{Validation of Random Sample}
\label{sec:app-random}

\begin{figure*}
    \centering
    \includegraphics[width=0.96\textwidth]{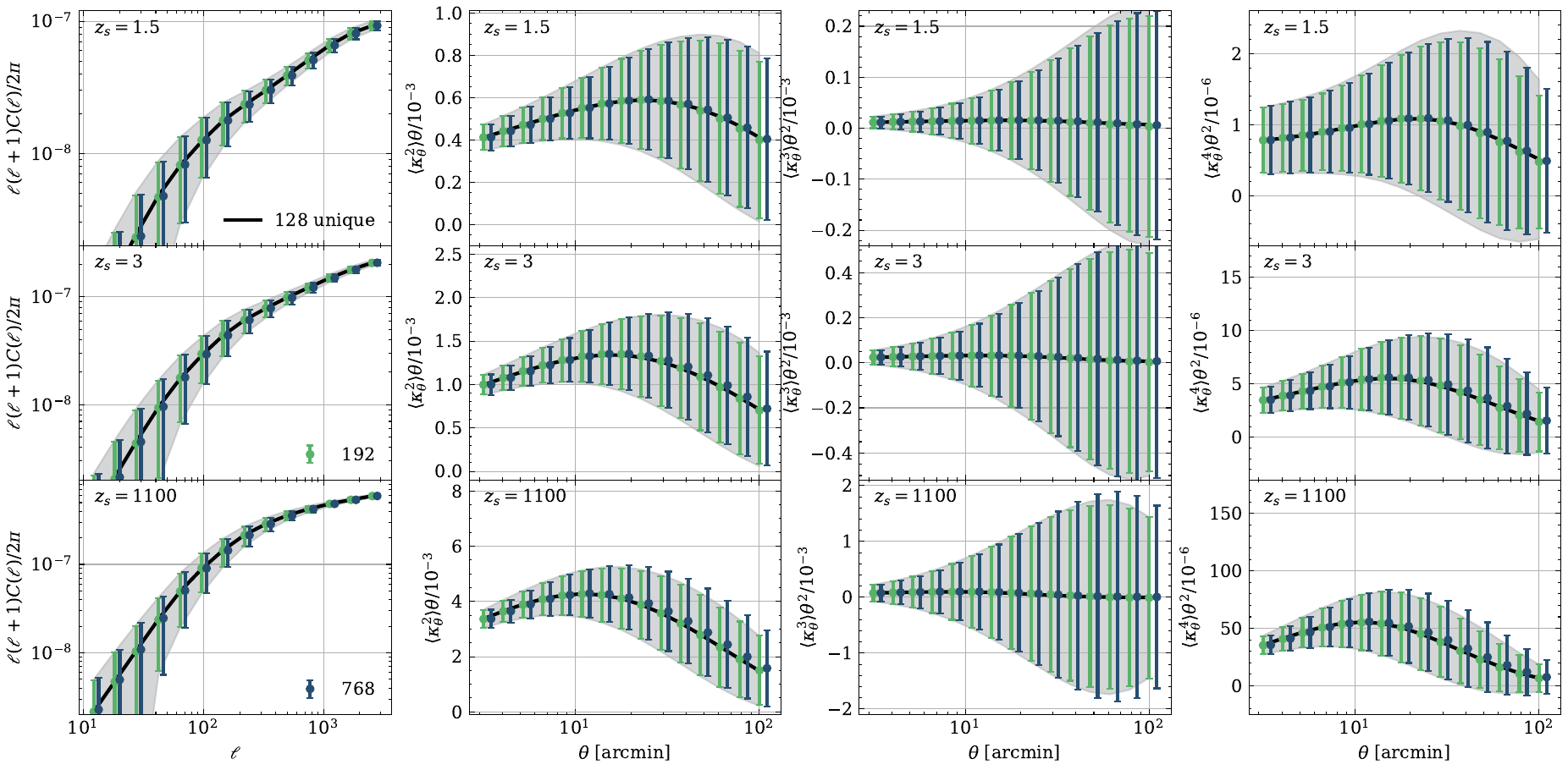}
    \caption{Convergence power spectrum and moments of three random samples.
    To enhance visualization, we multiplied the x-axis variables of the 768 samples (blue) by 1.1.
    The three random samples are statistically indistinguishable.}
    \label{fig:k234-L1200-random}
\end{figure*}

In Sec.~\ref{sec:results}, we treat the light-cones towards 192 random angles as the fiducial sample where the box replication effects are minimized, though some still suffer from repeated structures.
The reliability of this approach needs examination.
Fortunately, we can obtain 128 unique light-cones by implementing the pipeline with $N_\mathrm{side} = 8$ described in Sec.~\ref{sec:losf}.
Although the geometrical configuration of the HEALPix pixel differs from the mask utilized by us, the area of a single pixel, which is $53.7~\mathrm{deg^2}$, can cover the mask. 
Moreover, a more uniform sample is generated by splitting the whole sky into 768 directions.
Four statistics of three random samples are shown in Fig.~\ref{fig:k234-L1200-random}.
The statistical means and errors are all consistent, indicating that our fiducial choice is reliable.

\bsp	
\label{lastpage}
\end{document}